\documentclass[11pt]{article}

\usepackage{graphicx}
\usepackage{amssymb}
\usepackage{epstopdf}
\usepackage{setspace}
\usepackage{geometry} 

\def\x{\times}
\def\MI{{\em MELENCOLIA I}}
\def\MII{{\em MELENCOLIA I.1}}
\def\MIO{{\em MELENCOLIA I.0}}
\def\0#1{{\mathrm{#1}}}
\def\1#1{{\mathbb{#1}}}
\def\2#1{{\mathbf{#1}}}
\def\3#1{{\mathcal {#1}}}
\def\4#1{{\mathsf{#1}}}\def\5#1{{\mathfrak{#1}}}
\def\6#1{\overline{#1}}
\def\7#1{{\check{#1}}}
\def\8#1{{\widehat{#1}}}

\def\<{\langle}
\def\>{\rangle}

\def\x{\times}

\def\BE{\begin{equation}}
\def\EE{\end{equation}}
\def\BEA{\begin{eqnarray}}
\def\EEA{\end{eqnarray}}
\def\BI{\begin{itemize}}
\def\EI{\end{itemize}}
\def\BEN{\begin{enumerate}}
\def\EEN{\end{enumerate}}
\title{\Huge \MII\footnote%
{This corrects and replaces the 2006 version of arXiv:physics/0602185}
}

\author{David Ritz Finkelstein
\\
\copyright 2007}
\date{\today}
\begin{document}
\maketitle
%\vspace{6in}
%\begin{center}
%	 \begin{figure}
%	\vskip-1in
%	\hskip-1.5in\includegraphics[width=9.0in]{0MELENCOLIA}

%	\caption{\MIO,  A. D\"urer, 1514}
%	\label{fig:0MELENCOLIA}
%	\end{figure}
%	\end{center}

\abstract{D\"urer's engraving \MI\, was circulated in two versions not previously distinguished.  
Besides their conspicuous early Renaissance scientific instruments and tools, they contain numerous apparently unreported concealments whose detection reveals heresies expressed in the work. 
The main one is encoded in the motto {\em MELENCOLIA I} itself: 
Natural Philosophy, not Mathematical Philosophy or Theological Philosophy, is the way to knowledge.  
Unusual optical illusions and
subliminal images, differing between the two versions, declare the relativity and ambiguity of perception, and indicate that the work was a Humanist document intended for a Humanist viewership.}
\tableofcontents
\newpage

%\part{\Large \MI} % 0501219

\section{GATEWAY TO HEAVEN }
The engravings \MI\,  of Albrecht D\"urer  --- there are at least two versions extant ---
have  evoked and defied interpretation for centuries.
I offer an interpretation that
fits more of the picture with fewer assumptions than any other I have seen.
It is at least a new  way to enjoy this remarkable work of art.
I found it by the experimental method.
\MI\, is a world in itself, 
with enough  encryption and subliminal detail
to support the experimental method.
I inferred a theory of the symbols that I had seen,
used it to  predict the existence of further symbols,
and corrected it when its predictions  proved wrong.
This led me through some  wrong interpretations to one 
that seems to work as far as it goes.

\MI\,  (I will argue) is a 
public expression 
of a kind that  may arise when free speech 
is  suppressed but free thinking 
has not died out.
Under such conditions 
a private language can evolve of
ambiguous symbols 
having both overt and covert meanings.
The  overt meanings are innocuous.
The  covert 
ones are dangerous.
Were they not deniable they might be
capital crimes.

To understand 
how D\"urer meant
his viewers to take this work
without projecting views of our time
onto him,
I rely mainly on
sources demonstrably 
used by D\"urer in his body of works
and arguably used  in
\MI\,:
the Bible, 
the {\em De Occulta Philosophia} of Agrippa, 
 the {\em Hieroglyphica} of Horapollo,
 and works of D\"urer himself.
They are crucial for the exegesis at hand.

I  draw on earlier
studies of \MI also.

The art historian Erwin 
Panofsky 
saw the engraving as intensely autobiographical
and read its melancholy as  D\"urer's own
frustration at the gap between artistic and divine creation
\cite{PANOFSKY}. 
This interpretation
was the starting point for my own reading
and survives as D\"urer's cover story.

Frances Yates,
historian of the Elizabethan age,
recognized that \MI\, refers to the
form of the melancholia theory proposed
in the {\em Occult Philosophy} of Agrippa von Nettesheim
\cite{YATES1979}:
that melancholia is mere  depression in ordinary people 
but   
can inspire
creative fervor in the gifted.
This recognition and the reference to Agrippa  were indispensable.
But Yates then read the engraving 
 as a declaration of the harmony between microcosm 
and macrocosm: as above, so below.
This is a Hermetic doctrine of the Faustian magus and clashes 
in spirit and letter
with the rationalism, Humanism, and modesty
I find in \MI.
I will show that
D\"urer eliminates the  Hermetic 
above and below by eliminating the Celestial Sphere.

The art historian Patrick Doorly
supports  Panofsky's 
understanding and goes deeper \cite{DOORLY}.
He too sees 
that the melancholy is that of the artist who cannot define absolute beauty.
For him, \MI\, is an illustration for Plato's {\em Greater Hippias}\/, 
a dialogue on beauty,
and  the angel is the melancholy artist.
As I demonstrate,  
the winged boy is the artist, not the angel;
her job-description is not to define beauty,
and her mood is intentionally ambiguous.

Peter-Klaus Schuster,
art scholar and museum director,
surveys in depth how 
\MI\,  has been read 
in the last half millennium
 \cite{SCHUSTER}. 
He
finds it to be  D\"urer's Book of Virtues,
implicitly likening it to the {\em Golden Book of Virtues}
of the poet Friedrich von Spee (1591-1635).

I   look more closely at the engraving than these writers did.
For example, I see
two significantly distinct states
of the engraving in circulation
under the one name
and description, so far undifferentiated in any museum catalogs
I have seen.
I call them \MIO\, and \MII\
and refer to them collectively as \MI.
Of
 dozens of clues waiting
 in the fine structure of \MI,
 I notice just two
 in the current literature, those 
 I call 
First Ghost  \cite{SCHUSTER} 
and the Wave \cite{HECKSCHER}.
Probably I have overlooked some myself.
I also depart from earlier writers in 
systematically using the D\"urer coat-of-arms
and the three well-known volumes I have mentioned
as keys to \MI.

\MI\,
has subliminal, superliminal, and hidden elements for the 
amateur sleuth to find and connect.
The subliminal elements I fit my theory around are
the  Wave 
\cite{HECKSCHER, SCHUSTER}, 
numerous serpents,  and
five quartets of subliminal faces, 
including three with their
bodies attached.
The hidden elements I 
find include 
a key Anagram,
assorted rebuses, gematria,  and puns,
a Star of David, the Bellringer, and
 numerous ambiguous perspectives.
 (For brevity I capitalize the specific 
 objects in \MI.)
 The superliminal elements I interpret are
the Chimera,
Octahedron,
 Angel, 
Purse,
Keys,
and Compasses;
the Millstone and
Dog;
and
the Ladder,
House,
Magic Square,
Boy,
Scales,
Globe,
Moonbow,
 Comet, 
 Flourish,
Tools, 
 Nails,
Sun-dial, 
Hour-glass,
and  Flourish. 

\subsection{D\"urer the Relativist}
D\"urer is relativistic.
This is no
 anachronism,
 merely a recognition of the roots of relativity.
D\"urer was famous as mathematician as well as artist,
one of the creators of
descriptive geometry,  the theory of how
the same object in space looks from different points of view,
probably
the  first mathematical theory of relativity
in the broad sense.  
He is  a significant figure in the
history of 
the perspective technique,
which had been practiced by 
the Romans and Greeks  
and 
 suppressed by the Church as deceptive.
He
brought ``the secret of perspective''  (his term)
 from Italy to northern Europe.
Some say that his book on the subject
founded the mathematical theory
 of descriptive or projective geometry,
 though this theory developed 
over centuries
with growing rigor.

Perspective theory, special relativity, 
and general relativity  all
concern  light
and so share family resemblances.
The apparent foreshortening of limbs in perspective,
and the  apparent shortening of rulers 
and slowing of clocks when they
move relative to us
or approach masses,
are all relativistic effects  in the broad sense.
Where relativists
 today speak of ``light-cones,''
Leonardo and D\"urer   
spoke of  ``pyramids of vision'',
light-cones with time omitted, 
made up of light-paths  in the space of the artist
rather than in space-time.
Leonardo and D\"urer  held that in perception one such pyramid 
diverges from the point observed,
 like a future light cone, and another converges to the 
 the observing eye, like a past light cone.
 The perspective transformations of the graphic plane 
that we carry out 
when we shift our viewpoint,
and the transformations of rotation and acceleration
of modern special relativity, are 
both
represented by $2\x 2$ tables of numbers, 
real or complex  respectively, making
projective geometry   still a  useful playground
for relativists today.
D\"urer uses his deep understanding of perspective to 
construct several  ambiguous puzzle pictures within the picture,
presenting the relativity of perception 
as a major limit to human understanding.
Again, when D\"urer adopted  Leonardo's project for
a mathematical theory of absolute beauty,
he relativized Leonardo's concept
by specifying projective transformations relating
thin, medium, and plump standards of relative human beauty.

\subsection{The Triumphal Arch}
I claim with some confidence that
\MI\, is intensely  programmatic and even ideoglyphic
because 
his immediately preceding work,
 {\em The Triumphal Arch of  Maximilian I}\/,
 is demonstrably so,  and
we have
its text message,  its key, and its
ideoglyphic transcription. 
It evolved in four stages:

In around the fifth century, the Egyptian Horapollo Niliaci,
a grammarian from Phanebytis, under Theodosius II (408Ð450 AD),
rendered 189 hieroglyphs into plain Egyptian
 \cite
{HORAPOLLO}.
When a Greek translation of his work, {\em Hieroglyphica}\/, was 
found in the fifteenth century,
it was widely believed.
Centuries after D\"urer,
the Rosetta Stone showed that 
 Horapollo was usually wrong but not always.

 In 1512/1513
Willibald Pirckheimer (1470-1530), 
D\"urer's learned neighbor, best friend,
Humanist,  state councillor,  and translator of
Greek and Hebrew classics,
translated 
{\em Hieroglyphica}  from Greek into Latin.
D\"urer illustrated this translation and
it  was widely disseminated,
providing a fresh graphic vocabulary that 
influenced the visual arts of  Europe.

Then Pirckheimer  wrote a Latin eulogy of the Holy Roman Emperor Maximilian I, 
patron of scholars and artists and specifically of D\"urer,
who illustrated writings of Maximilian himself.

Finally, using
the {\em Hieroglyphica} as dictionary,
D\"urer translated Pirckheimer's eulogy  word by word 
into glyphs like
dogs, goats, harpies, cocks, snakes, and scepters in
the centerpiece of the {\em Triumphal Arch}. 
Since  {\em Hieroglyphica} states that  the ancient Egyptians 
represent a king by
a dog with an ermine stole,
D\"urer drew the Emperor as a dog with a stole.
This dog is not a dog, it is a glyph for Maximilian I.
The text, key, and wood-cut of the {\em Triumphal Arch}
all survive.

We do not have the clear text for \MI\,
except that  
D\"urer tells us that the Purse and Keys represent wealth and power.
As a result many elements in \MI\, are overlooked
and
some that are seen are not read but taken as mistakes.
Some suggest that D\"urer could not
spell his motto, for indeed its
spelling is strange;
that he could not draw a cube in perspective,
since the Octahedron is not quite a truncated  cube;
that he could not make clean corrections, 
and indeed there are blurs and traces of another numeral
in the Magic Square.
There are  indeed well-known limits 
to D{\"u}rer's knowledge. 
Most famously,  the astronomer Kepler pointed out that
D\"urer 
drew the ellipse that is produced when a plane cuts a cone 
as egg-shaped,
with the end nearer the tip of the cone 
smaller than the other end
\cite{DURER}.
There is also  an error in optics in 
\MI.
More to the point, D\"urer 
declared that there are
ultimate limits to all human knowledge,
including his own,
and demonstrates this in the engraving itself.

Nevertheless,  to catch his meanings  
one should begin by recognizing his 
power with the burin,
his mathematical skill in perspective,
his dedication to precise language, 
and  his acute observation of nature.
The peculiar spelling of \MI,  the
odd shape of the Octahedron, 
and the curious anomalies in the Magic Square
are then meaningful.

\subsection{Clues}

Here are clues I found indispensable.

In 1453 Gutenberg invented printing with movable type and
Muhammad II, founder of the Ottoman empire,
 captured Constantinople.
 In 1455
he
invaded
Hungary and
Albrecht D\"urer's father,
a goldsmith of the same name,
23, left 
his village of  Ajt\`as (or Eytas) in Hungary,
 resettling in Nuremberg.
{\em Ajto} is Hungarian for door, 
{\it Th\"ur}  in the German of the time, 
so
Albrecht of Ajt\`as became 
Albrecht Th{\"u}rer,
roughly ``Albert the  Door-er".
{\em Th\"urer} can  mean door-maker, but as a name it means
``of Door" (Ajto) 
in the way that Frankfurter means ``of Frankfurt".
Had he moved to England
he 
might have 
become Albert Gates.
 He soon simplified the spelling to D{\"u}rer, a
  homonym in one dialect.
He married his master's daughter, Barbara Holper,
and they had eighteen children.
Only three lived,  the brothers Endres (Andrew),  Albrecht, and Ajt\`as
or Hanns.

Like Leonardo, 
Albrecht D\"urer the younger began his training under a goldsmith:
 his father.
By 1514,
the year on our engraving,
 he had made two
historic visits to Venice,
where  he studied 
perspective
and human proportion
and
developed 
a passion for geometry.
During D\"urer's  first visit to Venice (1505-1507),
Leonardo  finished the  {\em Mona Lisa}  (1503-1506).
He worked on her mainly in Florence, but
took her with him wherever he traveled, and 
never parted with her.
There are signs of 
her  influence in  \MI\,  so it seems likely that D\"urer saw her;
and in that case D\"urer and da Vinci probably met.
There is strong
Leonardo influence in other work of D\"urer.

D\"urer
brings us back to many other beginnings besides relativity.
The
art critic had not yet been invented.
The first published art criticism 
was  the praise of 
this very engraving 
by a younger friend, 
 Joachim Camerarius (1500-1574),
 who
watched D{\"u}rer  at work, 
 translated his book on human proportions
  from German into Latin, 
 and at the end wrote his eulogy,
declaring that 
although D\"urer's art was great, it was
the least of his accomplishments.
D\"urer may have invented the modern artist,
by
charging for his own  gift,
instead of  billing merely for time and materials like
a house-painter, as his contemporary Raphael still did.
He produced the first printed  map of the world as a 
sphere viewed from space,
and
the first printed star-chart, as
the celestial sphere viewed from inside.
He was  the first to publish 
a mathematical proof  or a book on pure mathematics
in the German language instead of Latin.
He shaped the German scientific language
much as 
Leonardo did the Italian:
by using the ordinary language
instead of flowery Latin, 
cutting out the prose 
ornament  and classical erudition that pervaded
scholarly writing in  his time.

He was a brilliant and recognized originator in art too. 
He made the first known self-portrait
and the first known
specific landscape.
He invented etching 
and 
perspective drawing machines.

D\"urer  gave graphic art a sanctity
above artisanry and comparable to
that then 
attributed to music, geometry, arithmetic, and astrology,
the medieval quadrivium.
Number and measurement lifted these
from the mundane  into  the divine,
so
Leonardo, D\"urer, and others sought to found art
and beauty
too on number and measurement.
The concept of the Golden Rectangle, 
Leonardo's famous drawing of a man inscribed in a circle,
and D\"urer's books on
descriptive geometry
and body proportions
are mementos of that quest.
So is much of \MI.

D\"urer
 was a Humanist in his interests in mathematics,  science,
 poetry, and antiquity
and moved in Humanist circles.
He argued philosophy hotly and on equal terms with the
famous Humanist scholar l
Pirckheimer, and conversed with the Humanist Erasmus.
He passionately  criticized corrupt ecclesiastical
practices and  supported the Reformation.

\subsection{Crime} %051015

Why all the mystery?
D\"urer could have explained his engraving for us.
Leonardo made up many riddles, 
and he wrote the solutions next to them,
protected  only by his mirror-writing.
 D\"urer and Pirckheimer left us the key to the 
 {\em Triumphal Arch}; why  
not  to \MI?
In his essay on this picture, furthermore,
Camerarius  
ignored its many concealments, 
described the Magic Square 
as a spider-web full of dead flies,
and omitted all mention of
the religious and scientific contents
\cite
{HECKSCHER}.
Why was he so inaccurate?

First, I must admit that 
Camerarius was not as wildly inaccurate as I thought at first.
Some  of the numerals 
in the Magic Square
are  z\"oomorphic,
 though 
not dead but alive, and
not flies but snakes.

I suppose that D\"urer and Camererius  were evasive in order 
to evade
 the  pyre.
In their times religious
thought-police persecuted
both the old magic and the  new sciences
with increasing ferocity.
Accusations of black magic and witchcraft beset 
both.
There was ample reason for
D\"urer and  his heirs
to whisper their dissent.
Yet the market for this engraving wore out its copper plate.
There must have been many eager and able to 
read his hidden message.

\MI\, was not D\"urer's first foray into dissent.
In about 1498 D\"urer  published 
fourteen woodcuts of the {\em Apocalypse}\/. 
One of them, 
{\em The Opening of the  Fifth and Sixth Seals}\/,
shows the wicked hiding from  God's vengeance.
The wicked are the Pope, Cardinal, Bishop, Kaiser, 
and Kaiserin, 
the top of the hierarchy of the Catholic Church 
and the Holy Roman Empire.
Another,  {\em The Battle of the Angels}\/, 
shows the imminent execution of the Pope by an avenging angel.
So by the age of 27,
16 years before the date of \MI,
D\"urer --- with his master Wohlgemut --- 
had already testified against the hierarchy of the Church
and publicly killed the Pope and the Emperor in effigy.
On the one hand, this
verifies that D\"urer had the courage to express
dangerous views.
On the other hand,  it suggests that
the views lurking in this engraving
are still more dangerous, since he hid them.
We  can also infer that
these views  would incite  Protestant censors,
since D\"urer had already flaunted Catholic law.
What was the crime of Albrecht D\"urer?

The crime begins to take form when
I introduce an accomplice.

\subsection{Chimera}%050404

A Chimera
with  head of mouse,  wings of  bat,
and tail of serpent flies at us out of the engraving.
It carries the Banderole with the Motto
``MELENCOLIA{\S}I" that gives the engraving its name.

%	\begin{figure}
%	\vspace{1in}
%	\begin{center}[In preperation]
%	%\includegraphics[width=2.0in]{0BAT}
%	\caption{Chimera. \MI\,  detail.}
%	and  D\"urer
%	\label{fig:0BAT}
%	\end{center}
%	\end{figure}

The Motto  refers
to the ancient Greek humor of melancholia, 
and specifically to a  revolutionary
form of the humor theory, derived from
the teaching of the Florentine Neo-Platonist Marsilio Ficino (1433-1499) 
by Heinrich Cornelius Agrippa von Nettesheim 
(1486-1535),
a prolific and controversial figure in the Florentine 
Neo-Platonist school.
The theory of melancholia, the
Magic Square,  and much of the cosmology
in this engraving may have come to D{\"u}rer directly
from Agrippa.

\subsection{Agrippa of Nettesheim}
\label{sec:AGRIPPA}
Agrippa never settled in one city long,
but wandered Europe, seeking royal support, 
making enemies
wherever he went and fleeing them, 
leaving behind centers of intellectual
activity \cite{NAUERT}.
He  visited Nuremberg in 1510 with 
the manuscript of his {\em De Occulta Philosophia}\/,
an encyclopedic compendium of astrology, alchemy,  Cabala,
 and many propositions of biology, physics, and medicine,
 mostly false;
he wrote that diamond demagnetizes magnetite, for one example among many
 \cite{AGRIPPA}.
He said that he studied the occult philosophy
 because he believed that it traced back to
 a secret divine revelation given
 at the same time as the public divine revelation of the
 Torah.
 
The engraving is based on the Neo-Platonic cosmology of the time,
which had three Worlds, explicitly numbered  by Agrippa:
\BEN
\item  World I is the Terrestrial or Elemental World,
our material  abode,
made of the four elements distributed in space and time.
\item  World II is the
Celestial World above,  with one rigidly rotating  sphere for the fixed stars
and many others for the planets.
Ptolemy used over 90 spheres in 
his Celestial World.
\item World III is the Spiritual or
Intellectual World outside the Celestial World,
where Angels and Platonic Ideas live.
\EEN
Neo-Platonists divided the human 
psyche  into three  corresponding Mentalities or Faculties specially 
suited to these Worlds
\cite{YATES1979}.
 \BEN
\item {\em  Mens imaginatio}\/, the Imaginative Faculty,
 empowers the artisan, the artist, and the Natural Philosopher
  to view  arrangements of the 
four Elements in Space and Time, but nothing more.  

\item {\em Mens ratio}\/, the Rational Faculty,
allows astronomers and Mathematical Philosophers
to know when stars rise and set and
thus through astrology it allows
 statesmen to know when  kingdoms rise and fall.
 
\item {\em Mens contemplatrix}\/, the Contemplative Faculty, 
enables Theological Philosophers to know  angels and Plato's Forms or Ideas. 
\EEN

The names for the three Worlds vary.
Agrippa also called them
{\em idolus} or {\em imaginatio}, {\em ratio}, and 
{\em mens} respectively
\cite{NAUERT}.

Agrippa also offered three increasingly reliable 
Gateways to knowledge of God, through three Books and three Philosophies,
belonging to the three Worlds:
\BEN
\item Gateway I is Natural Philosophy, the study of the Book of Nature.
\item Gateway II is Mathematical Philosophy, the study of the Book of Law, the Hebrew Bible.
\item Gateway III is  Theological Philosophy, the study of
the Book of the Gospel.
\EEN
In Book I we see merely the creatures of God, not God,
Agrippa wrote.
We may use Reason to deduce the existence and nature of God,
but original sin clouds our souls 
and makes our reason unreliable.
But Books II and III are the result of revelation
and require only
 faith.

Ficino and Agrippa also kept the ancient Greek 
psychology of the four humors,
one
for each Element:
blood for Air and the sanguinary temperament, 
yellow bile or choler
for Fire
and the bilious temperament, 
phlegm for Water and the phlegmatic temperament, 
and black bile or {\em melan cholia} for Earth and acute depression.
This theory occupied D\"urer's imagination in 1514 \cite{PANOFSKY}.
It too survives today,  in our vocabulary for temperaments.

Melancholia, the black bile, is a Greek myth, invented 
to fit the mental illness  of acute depression 
into humor theory.
Sometimes it pays to postulate previously unseen entities;
we found Uranus and the neutrino that way.
Sometimes it doesn't; there isn't any black bile.
Some
said  the appendix secreted it,
some the spleen.
Neither secrete anything.

Astrologers had associated Saturn, the slowest and darkest planet, 
with  the darkest temperament,
melancholia;
with the lowest Element,  Earth;
and  with the lowest people, 
taken to be those who work 
on or in the earth.
Agrippa
promoted Saturn from the lowest status among the planets to the highest, arguing that
Saturn is highest from the Sun and most sedate of all.
He attributed to Saturn the power
to inspire all three 
Faculties to their peak creativity, 
and transferred the demi-godlike geometers --- ``earth measurers,''
after all --- to the control of Saturn, with the farmers and miners.
According to Agrippa,  
Saturn could inspire the Imaginative Faculty of the artist with a form of melancholia that appears as
a creative frenzy or furor.
Agrippa also wrote a tract on the natural superiority of women 
over men.
On his part, D\"urer aimed to promote art and artists,
who were assigned to the Elemental World of Saturn,
from the lowest to the highest rank.
This may have made them natural allies,
at least for a time.
Later Agrippa
was  linked with the devil by Martin Luther (1483-1546).

He  remained in
the Catholic Church,
adopted an uncritical fideism,
abjured astrology,
and doubted all access to truth
but faith in the Bible.
 His work 
on the uncertainty and vanity of the sciences,
exalting faith in Scriptures and divine revelation
over all the arts and sciences, including astrology,
was written after his{\em Occult Philosophy} 
and rejected it, 
but appeared in print earlier.

Panofsky and Yates saw many Saturnian melancholic
elements in the engraving, in either 
the more depressive sense of Ficino or the
more creative sense of Agrippa.
They take it that D\"urer adopted
Agrippa's esoteric philosophy
 at least for the purpose of this engraving.
 \MI\,  converses
with that philosophy
but it also expresses some of Agrippa's later pessimism and rejection of astrology,
rejects his fideism as well,
and is more radically Humanistic  than the
writings of Agrippa 
or Luther.
 While Erasmus, Agrippa
 and D\"urer all took the  Christian Bible
 as divine revelation,
 they differed with 
respect to the Reformation.
Erasmus remained loyal to the Roman Church,
and distanced himself from Agrippa
when Agrippa's protest against the corruption 
of the monks who ran the Inquisition became 
radical \cite{GELLERD}.
D\"urer left the Roman Church to follow Luther.

D\"urer probably rejected astrology before Agrippa.
In 1514 Copernicus (1473-1543) 
had not yet published his heliocentric cosmology, but 
the cardinal, mathematician, 
and philosopher Cardinal Nicholas of Cusa (1401-1464) 
had long demolished 
the Ptolemaic spheres. 
Leonardo
had already written that the Earth moves, not the Sun, and 
that ``those who have chosen to worship men as gods
-- as Jove, Saturn, Mars and the like --
have 
fallen into the gravest error." 
He had derided necromancers
and excluded every trace of 
astrology from his intensely rational works on astronomy. 
Similarly, a 1494 picture of the folly of `` Attention to the Stars''
plausibly attributed to D\"urer 
in the {\em Narrenschyff} of Brant
\cite{BRANT}
shows
the astrologer-fool immersed in a motley flock of
misshapen fowl flying in random directions,
presumably
representing the products of 
the astrologer's disordered mind.

Martin Luther nailed his theses
to the church door  in 1517,  
and D{\"u}rer  soon became an ardent follower.
Luther specifically 
linked Agrippa's {\em Occult 
Philosophy} with witchcraft,
and Agrippa 
aroused his ire when he defended a
woman  accused of witchcraft
from her corrupt accusers.
Clearly D\"urer and Agrippa looked in opposite directions,
one   to the future and the other
to the past.

The motto labels the Chimera as Mr. Melancholia,
hence as
Agrippa,
who had recently passed through Nuremberg 
with his {\em Occult Philosophy} in manuscript form.
In view of the motto,
the darkness of the engraving 
can be  taken to indicate Agrippan
 melancholia, as Yates says;
 not depression but creative frenzy.
 Agrippa was such a charismatic figure
 that D\"urer's audience may have known  at once
 who he meant by the Chimera 
 carrying Agrippa's famous catch-word.

 Some say that the motto names the Angel diagonally 
opposite it.
In his
other works D\"urer  
puts his labels on or near the things labeled.
I suggest that this motto not only 
identifies the Chimera as Agrippa but also conceals a message,
and return it later.

If D\"urer  shared many
Agrippan teachings,
at least for this engraving,
it seemed excluded that he would
represent  Agrippa as a bat, let alone one with a serpent's tail.
The  theory hung in the balance until I looked up the bat in
the {\em Hieroglyphica}\/.
D\"urer made Agrippa bat-like
for the same reason he made his emperor Maximilian I
a dog: 
The {\em Hieroglyphica}
told him to:

``When they wish to indicate a weak and rash man, they draw a bat. 
For the bat, though it has no feathers, flies."
\cite{HORAPOLLO}.

Agrippa might seem weak to D\"urer simply for remaining in the Roman Church and abandoning both Natural Philosophy and Mathematical Philosophy.
He might equally well seem rash. In 1531 a plague broke out in Antwerp.
All the doctors fled but Agrippa, who stayed to tend the sick.
When the doctors returned they filed charges against Agrippa for practicing medicine without a license,
to recoup their wealthy patients, and Agrippa fled.

As for the serpent's tail on the Chimera:
In the {\em Hieroglyphica} 
a serpent represents the king (again), 
but:``When they wish to symbolize a king ruling
not the whole world but a part of it, they draw a serpent cut in half.''

So the bat/mouse/serpent symbol need not demonize Agrippa 
but may simply describe him candidly. 
Viewers of the time who knew their Horapollo and their Agrippa
may have 
recognized the Chimera and Banderole  as capturing the very essence of Agrippa.
Moreover if D\"urer were accused of
glorifying Agrippa, he could claim to have rather ridiculed  him.

In what follows
I assign each of the symbols I see in \MI\, 
to
one of the three Worlds of the Agrippan cosmology.

\section{THE THIRD WORLD}

\subsection{Four Ghosts} %060228
\label{sec:GHOSTS}

I was first drawn to examine \MI\, in detail when I learned 
by chance (from a briefly glimpsed internet site) of a subliminal face
 that I had overlooked for decades 
Step back several paces from a good print and look
at the shading of the front face of the Octahedron 
for a minute or so,
and you too will probably see a woman's face in profile.
This can be a {\em pons asinorum} for \MI:
You must see this face in the Octahedron
before you can hope to see the subtler
images in this engraving.

One calls such images subliminal
because they wait just below the {\em limen} or threshold of perception.
Here  the term subliminal itself is a subliminal reference to D\"urer.
The root {\em limen} can mean threshold, lintel,  gate, or limit.
Gates and doors dominate the interpretation of \MI.
The name D\"urer itself can mean ``door maker'' or ``man of the limen".

Schuster sees
 this face as a frightening death-head \cite{SCHUSTER}
and I saw it as a loving 
remembrance of D\"urer's mother, who died in the year 
of the dateline.
Likewise, some see D\"urer's last portrait of his mother 
as malicious, while some see it as an accurate record
of the physical consequences of 
bearing 18 children.
The emotional responses to the ghost of Hamlet's father also varied
from character to character.

I eventually noticed four subliminal faces in the upper 
pentagon exhibited to us by the Octahedron.
I  call them the four Ghosts. 
Which we see depends on how we
hold our heads relative to the print,
for our remarkable propensity to find faces in 
hiding is tuned especially to upright faces.

 $\bullet$ First Ghost, well known,  is the woman in profile with her head erect, already mentioned.

$\bullet$ Second Ghost  
 is a man with his head cocked $30^{\circ}$ clockwise. 
 
$\bullet$ Third Ghost appears when we stand the picture on its left edge.
At the left edge of the face of Second Ghost is a vee-shaped
dimple in the stone that becomes the nose of Third Ghost.

$\bullet$  Fourth Ghost is a younger bearded man seen in full profile looking to our left with his head upright,
to the left of First Ghost.

I found  Third and Fourth Ghosts
 some time after First Ghost and Second, but
not by careful search.
I inferred them on grounds
described below
and then found  them at once.
The poses of First Ghost and Second Ghost are 
those of D\"urer's last portraits of his mother
and father and
so perhaps they fit  into Panofsky's view that
the engraving is autobiographical.
Then Third Ghost might be D\"urer the Son.
In his last portrait of his mother,
D\"urer shows her 
divergent 
strabismus, a family trait that he inherited
known as the  {\em D\"urerblick}\/.
 Third Ghost  seems to be wall-eyed.

We have to scan for these subliminal  faces,
also called secondary images,
 with some persistence
before they emerge. 
They 
are subtle, ambiguous,
 and strain our perception;
 especially Third Ghost.
Do we see them or create them?
Most of the left edge of the face of First Ghost
has almost no physical correlate
in the engraving.
D\"urer compels  us to see it, when we look for it,
 by supplying chin and forehead lines
and eye and nostril shadows, but
we ourselves create the rest of the face.
Part of the art  
behind the subliminal  faces is that they 
are ambiguous, lie at the edge of perception,
and are eminently deniable.
They are
joint productions of the hand of the artist
and the mind of the viewer,  demonstrating
D\"urer's mastery of both media.
When we see one of the faces in this quaternity
we  do not see the other three, mainly
because of  elements they share.
For example,
 Second Ghost and First Ghost  share an eye-pit.

We are remarkably apt at
finding faces where there really are none.
Some can see a face in the Moon;
I see several in  maps of the ocean floor.

I have diligently tried to suppose that the 
subliminal faces in \MI\,
are products of my imagination.
D\"urer generously made this difficult
by organizing them
into quaternities.
Subliminal faces are common
in his work and that of others, but
scanning,
digitization, or lithography creates accidental ones
as well as distorting or destroying
those already there.
In an earlier version of this 
analysis,
drawn mainly from lithographs in books,
 I claimed that in D\"urer's drawing 
{\em Orfe der erst Pusaner}
there was a woman's
face in the tree top.  
In a finer reproduction
the lady vanishes, so I base
the present analysis on laser prints
with no perceptible grain.
I recognized the organization into quaternities 
only after finding several subliminal faces.
There are famous ambivalent faces,
like the one called ``The Lady and the Hag." 
Penrose made a trivalent optical illusion,
his Impossible Triangle, any one corner of which can be seen as a right angle.  
Quadrivalent illusions, however,
covert or  overt, occur in no other work of art
that I know of. 
Accidental subliminal faces happen;
accidental quaternities do not seem to. 

D\"urer's five subliminal quaternities
 go beyond
the familiar kind of relativity, 
the change in visual appearance under rotation
that turns $\uparrow$ into $\downarrow$.
Each face appears and disappears as
we merely rotate it about the line of sight;
to see one at all we have to be in the right 
reference frame because
our innate face-detector is most sensitive by far to 
upright faces.

The likeness of the First and Second Ghost to D\"urer's
parents is not convincing.
It seems likely that the faces are there not to commemorate
but to demonstrate,
namely that perception is not reliable.

D\"urer hides the four Ghosts  as
paper-makers hide watermarks.
Our perception adjusts itself to the largest range of 
brightness that is present.
We filter out distracting low-contrast background in
brightness within a high-contrast image.

Jan Wierix (1549-1615), ``the Flemish D\"urer'',
 made from scratch (so to speak)
 an engraving that is also called \MI\,  and is
 usually described as a copy
\cite{FAMSF}.
Wierix actually retained almost all the superliminal content
but revised the subliminal content substantially.
In particular, Wierix eliminated at least two faces of every
quaternity
and  substantially changed some of  the surviving faces
and other details, in a way that
erased the subliminal message of
D\"urer's work.
His erasure seems too systematic to be chance.

Viewers of D\"urer's time may well have 
seen 
these faces easily and quickly,
 alerted  by the many 
subliminal faces in earlier works of D\"urer. 
We mention some as precedents
and to provide practice in face detection.

The most spectacular 
demonstration of D\"urer's wizardry with illusion is
his engraving called {\em Der Spaziergang}\/,
or  {\em The Lady and the Gentleman}\/, 
or  {\em Young Couple Threatened by Death}\/.

$\bullet$ A brilliant subliminal image of
a male face
forms out of and looms over the {\em Young Couple}\/. 

Their heads form his eyes,
 her arm his mouth.
The gentleman's plumed hat becomes a flaring
eyebrow of the subliminal face.
  In this picture D\"urer does not use the watermark 
  concealment
  technique but a kind of bricolage.
Several overt forms have to be broken into pieces
by our perception and reassembled  to make
 the covert one.
 
 $\bullet$ His watercolor {\em View of Arco} has a cliff that scowls
famously. 

$\bullet$
His drawing of the same
Arco hides three such angry faces
in the landscape.

\MI\, is one of three works called the Master Engravings  of D{\"u}rer,
all dated within a year of one another.
The other two  are   {\em The Knight, 
Death, and the Devil} and  {\em St. Jerome in His Study}\/. 
D\"urer portrays the Knight as spectacularly oblivious to
 death and the 
devil, though in all other such encounters in the work of D{\"u}rer,
  the knight engages evil in mortal combat. 
The Knight is generally said
to represent the ideal Christian Soldier described by 
Erasmus
 (1466?-1536),
 so virtuous  that he was blind to the evil about him.
Erasmus was a revered 
acquaintance of D{\"u}rer. 

On the other hand,
D{\"u}rer wrote a fervent plea 
to Erasmus,  perhaps unsent, 
to join with him and Luther in battle against the ``Cave of Hell''
and die a martyr's death.
Erasmus  remained 
within the Catholic church and worked for reform 
instead of  applying for martyrdom.
The Christian Soldier and the unseeing Knight 
may both have represented  Erasmus himself for D\"urer.

In any case D{\"u}rer pays the Knight a subtle  tribute.
Others have pointed out one of the subliminal faces
 in {\em The Knight}\/.

$\bullet$ Part of the garment of Death forms 
a secondary image of an agonized face 
smashed between the 
Knight's two fists.

The Knight
may seem blind  to Death
but still attacks him.

$\bullet$ A second subliminal face watches the Knight
 from the rocks of the hillside.

Eventually,
Pirckheimer tells us,
 the high-minded D{\"u}rer would 
lament that the behavior of the Protestant clerics
 made even the Catholic clergy look respectable by comparison,
 and there is indication that D\"urer never lost his
 high regard for Erasmus:

$\bullet$
  D\"urer's 1526 portrait
  {\em Erasmus of Rotterdam} 
has a gleeful subliminal face  high on the left sleeve of Erasmus.

These examples show 
that D\"urer's subliminal faces are not just virtuosity or puzzles
 and games.
They advance the story
 like a Greek chorus.
They often tell us
how D\"urer felt about the subject.
For example, I
suppose that the cliffs of Arco scowled
because the Humanist
D\"urer disapproved of practices 
connected with the Arco cathedral,
a vivid graphic use of the pathetic fallacy.
The {\em Young Couple}  suggests
D\"urer and his 15-year-old bride Agnes 
soon after their marriage, just before
D\"urer left plague-stricken Nuremberg and his new bride for Italy, where 
he met with their 
neighbor Pirckheimer.
Then Death hiding behind the tree
is the plague,
and the bulbous-nosed beetle-browed 
figure that unites
the {\em Couple} could be the
powerful Pirckheimer.

%	\begin{figure}
%	\begin{center}[In preperation]
%	\vspace{1in}
%	\caption{``The Lady and the Gentleman". D\"urer, MORE}
%	\label{fig:LADY}
%	\end{center}
%	\end{figure}

\subsection{Octahedron} %050513

D{\"u}rer seems to put his Octahedron 
--- 
the name given by
\cite{FEDERICO}  ---
in a place of honor. 
The artist stands before it and
it balances the Angel in the composition.
Yet it is  bleak and sterile against the Angel. 
One looks for a deeper 
significance to justify it artistically,
or at least to understand its shape.

%	\begin{figure}
%	\vspace{1in}
%	\begin{center}[In preperation]
%	%\includegraphics[width=2.0in]{0OCTAHEDRON1}
%	%\includegraphics[width=2.0in]{0OCTAHEDRON2}
%	%\includegraphics[width=2.0in]{0OCTAHEDRON3}
%	\caption{Octahedron in three views. \MI\, detail.}
%	\label{fig:0OCTAHEDRON}
%	\end{center}
%	\end{figure}

A rhombus is a  planar quadrilateral whose sides are all equal. 
A rhomboid  is a closed surface formed of six congruent rhombuses.  
All the descriptions we have encountered
 see the Octahedron as a truncation of a rhomboid with its longest diagonal  vertical \cite{MATHWORLD}.
 
 If it represents any  of the Neo-Platonic Worlds,
then its
mathematical nature 
and its altitude in the picture unambiguously
mark it as the Intellectual World III,
the abode of Platonic Ideas and angels.
The Ghosts in 
the Octahedron also support this conclusion, 
which puts them in a philosophical Heaven.
I adopt this interpretation in what follows.

If the two missing vertices of the Octahedron were restored, 
they would seem to  lie in a vertical line in space.
The compulsion to see the figure as a symmetrically truncated rhomboid
is strong. 
It then has a vertical axis of three-fold symmetry.

In principle, however,
 one cannot determine the form of a solid from one perspective view.
 One can shift any surface point along the line of sight from the eye 
without changing its apparent location in the perspective view.
Therefore there are many solids all having the same perspective view.
The Reversing Staircase, the Necker Cube and the Cameo Intaglio 
exhibit this ambiguity.
The ambiguity of perspective is an important limit to our knowledge of the world.
For example, our view of the stars gives us no inkling of how far they are. 

Sometimes  other cues are so strong and unconscious 
 that we cannot help inferring
a particular solid.
For example,
it is hard {\em not} to see a perspective of a cube as  a cube,
or a perspective of a person as a person, even though intellectually we know
the picture is ambiguous.
The D\"urer Octahedron is subtler than the illusions we mentioned, in that they are merely 
ambivalent,
while the D\"urer Octahedron is at least trivalent.
View the Octahedron with head erect, and one sees   First Ghost
on a truncated rhomboid.

View it with head cocked to the right and the First Ghost,  rhomboid, and
 threefold symmetry axis disappear from our perception.
Instead one sees  
Second Ghost on a nearly rectangular slab with two 
diagonally opposite corners trimmed, cocked and cantilevered back toward the horizon.

Turn the engraving on its side to see Third Ghost and the Octahedron changes shape again.

I do not recall seeing another optical illusion that changes its shape when we turn it.

Some claim to measure the angles of the Octahedron from the engraving.
This is impossible.
These angles  depend  on implicit assumptions that  are not deducible from the engraving.
That impossibility is the point.

 If one compresses the engraving vertically by a factor of   
 $\sqrt{\varphi}$, 
its frame becomes nearly square, 
and the Octahedron could pass for a truncated cube; but only
until one actually juxtaposes a true cube, or a perspective of one,  for comparison.

One acceptable interpretation sees 
a truncated rhomboid with the apical  angle  close to $80^{\circ}$ \cite{MACGILLAVRAY}.
If one juxtaposes a model of such a rhomboid with the engraving,
the match is good \cite{KHULUSI}.

Before the engraving D\"urer  made what seems to be a rough sketch of the Octahedron that has survived, 
though it has been recognized as that only recently
\cite{WEITZEL}.
The sketch shows an irregular pentagon inscribed in a circle.  
   The apex angle in the drawing is indeed $79.5^{\circ}\pm .5^{\circ}$.
   This agrees  well enough with Macgillavray's interpretation.
   
   The sketch shows what seems to be a regular heptagon.
   The irregular pentagon is inscribed in the heptagon by omitting
   two vertices of seven, keeping one between them.
   There is no way to make a regular heptagon by 
   Euclid's methods, but 
   D\"urer gives a  simple 
 approximation in his posthumous 
   work on descriptive geometry.
 Theoretically, the sharpest angle of the inscribed pentagon
 is three fourteenths of a full circle, or about $77.1^{\circ}$. 
Perhaps  D\"urer merely copied his imprecise sketch, for the exact angle is unimportant for the illusion,
as long as it is close to a right angle but not too close.

Leonardo had 
already undertaken a mathematical theory of beauty.
That goal would occupy much of D{\"u}rer's 
later years, 
and result in his works on human proportions and on how to construct projections and 
perspectives with compass and ruler. 
D{\"u}rer's study of human proportions is an amazingly dry gallery 
of stark outline drawings of standing human nudes,  with tables of anatomical dimensions to three 
decimal places. 
The human measurements were to be the foundation of a geometry of beauty,
as stellar measurements were the foundations of Ptolemy's geometry of the heavens.
Music was already divine and part of the quadrivium because it was considered mathematical. 
By  providing a mathematical basis 
for art, D{\"u}rer hoped to sanctify the graphic arts as well. 
Possibly this belief in a mathematical theory of art 
was only one aspect of a belief in a general mathematical wisdom, a mathesis.

I propose that D\"urer  
designed the Octahedron to be
ambivalent, irresistibly construed
 as a truncated rhomboid in one orientation,
as a truncated slab in another, and as something else from yet another.
 For this reason he had to both stretch the cube and truncate it;
 neither alone would have created an ambivalent
 solid, for we are ready to see cubes in any orientation.
 The exact angles of the Octahedron are then immaterial as long as they are
 not too close to right angles, which might force themselves upon our perception
 in any orientation.
The Octahedron is a puzzle that is unsolvable in principle
but appears to solve itself
in some views.
It does not simply show off D{\"u}rer's 
mathematical muscle.
It declares that the Intellectual World may have a mathematical design, 
but if so that design is inaccessible to us.

\subsection{Angel} %0501015
\label{sec:ANGEL}

Angel-wings indicate sanctity,
eagle-wings fame.
These sacred wings brush and
bless 
the hour-glass, the scales, 
the numeral 1 in the D\"urer Table, 
and nothing else in the picture.
The Angel thus blesses the scientific instruments 
above the workman's tools that litter the ground. 
She holds a sealed 
book and an apparently idle pair of compasses.
She cannot be using the compasses
as compasses when she holds them as she does,
nor has she a suitable working surface.

Most commentators also say that she is melancholy, 
or even the spirit Melancholia 
herself. 
The motto on the banderole, 
the shadow on the Angel's face, and the 
fist on her cheek are indeed consistent with melancholia. 
But  the hidden  motto is not negative
in mood  but positive
and in any case is attached to the Chimera,
not  the Angel.
Her  Water-cress and Water-ranculus
protect her from Earthy melancholia
by their Watery nature \cite{PANOFSKY},
so she cannot be melancholy according to Agrippa himself.
The Angel's expression is alert and focused,  not soft and sad.
Above all her gaze is not downcast, as was 
mandatory for portraits of melancholia, but elevated,
and her lips are possibly smiling, possibly straight, but certainly not drooping.
%	\begin{figure}
%	\vspace{6in}
%	\begin{center}[In preperation]
%	%\includegraphics[width=6.0in]{0SMILE}
%	\caption{The Angel smiles or does not. 
%	\MII,   detail.}
%	\label{fig:0SMILE}
%	\end{center}
%	\end{figure}

In \MIO\, there is a 
definite upward line at the near edge of her lips
that can be read as a  smile.
The far edge of her lips is almost hidden 
by the turn of her head,
so that only a trace of the smile is visible there.
The whole meaning of the Angel's expression, of the Angel
as a whole, and of the
entire engraving is controlled by that almost imperceptible
trace of the burin,
creating a high-intensity focal point  of a kind found elsewhere in D\"urer's work.
In 
\MII\,  the Angel has a darker face and
the place where her lip turns up in  \MIO\, is solid black.
This does not mean that she is not smiling, but leaves it for
us to decide. 
 \MII\,  is more serious in tone
than  \MIO\, 
but also more  ambiguous.

The ambivalent smile of the Angel may  be 
D\"urer's version
of the ambivalent smile of {\em La Gioconda}.
Leonardo finished the {\em Mona Lisa}\/, 
presumably in Florence, while D\"urer
lived  in Venice.
The Mona Lisa smile is wry. 
Her lips curl upward significantly more on her left side
than on her
right,
so that the viewer oscillates restlessly between two interpretations.
This is sufficiently unusual in art so that
when D\"urer's angel too has an ambiguous expression 
it is fair to look for influence.
It is easy to find once we look for it.
The bird's-eye view of background landscape 
in the {\em Mona Lisa}
was an
innovation of Leonardo,
demonstrating his mastery of perspective.
D\"urer 
adopts such a perspective  for the background of
\MI.
There is no direct 
evidence that D\"urer ever saw the {\em Mona Lisa}
but neither is it excluded, and
others have considered it likely that D\"urer saw works of Leonardo.
Since he had traveled for a month from Nuremberg across the Alps to Venice
  to study Italian art for over a year,
it would have been unnatural not to cross the smaller mountain 
between Venice and Florence 
to see the great Leonardo.
The engraving supports this supposition 
with its distinctive facial expression and background.

The two expressions differ significantly, however.
We do not doubt that La Gioconda smiles
with at least half her lips.
Only what she feels puzzles us. 
But we can never  see clearly
whether the Angel is smiling or not.
If she moved her sleeve by a millimeter or turned her head by a degree
the ambiguity would dissolve.
D\"urer posed her so that 
the Angel's expression could be taken as serious
or smiling depending on the expectation of the viewer.
This is not a picture of a facial expression
as much as a picture
of ambiguity itself.

In no case is the Angel melancholy, 
since she is looking up.
The pose  shows contemplation, 
as in Rodin's {\em Thinker}, not
melancholy. 
This and her wings say that she is the Contemplative Faculty
proper to the Intellectual World of Gateway III, 
Theological Philosophy.
Let us see how this interpretation fits the rest of the engraving,
and why Theological Philosophy is smiling, if she is.

\subsection{Fools}

This is a curiously domestic Angel.
She is dressed as a housewife, complete with keys and purse
hanging from her belt
which D\"urer says represent power and wealth,
and she attracts Christians and non-Christians alike.
Only a fool would worship her, 
the artist seems to say.
And there the fool is,   a crude low-browed 
subliminal caricature 
 lurking near her feet in the hem of her gown,   
 next to D\"urer's monogram
 Call him First Fool.

$\bullet$
 We must turn the engraving about 
 $60^{\circ}$ 
 counterclockwise to see First Fool best and then he looks to our right.
His nose forms part of  the bottom edge of the Angel's robe.
He has a mustache and curly hair.
%	\begin{figure}
%	\vspace{1in}
%	\begin{center}[In preperation]
%	%\includegraphics[width=1.0in]{0FOOL1}
%	%\includegraphics[width=1.0in]{0FOOL2}
%	%\includegraphics[width=1.0in]{0FOOL3}
%	%\includegraphics[width=1.0in]{0FOOL4}
%	\caption{Four Fools. 
%	``MELENCOLIA I",  detail}
%	\label{fig:FOOLS}
%	\end{center}
%	\end{figure}
A powerful winged housewife 
 with  a  fool in her hem also resides
in a woodcut 
 attributed to D\"urer
 in Chapter 13, On Amours ({\em Von Buolschaft}))
 in the {\em Ship of Fools} of Brant \cite{BRANT}.
 There
the winged housewife  is Venus, 
the goddess of 
the fools of love, 
the most common variety of fool according to 
Brant.
. 
%	\begin{figure}
%	\vspace{1in}
%	\begin{center}[In preperation]
%	%\includegraphics[width=4.0in]{0VENUS}
%	\caption{Frau Venus. 
%	\cite{BRANT}, Chapter 13, On Amours ({\em Von Buolschaft};
%	``On Courtship" is close to the root meaning.)}
%	\label{fig:VENUS}
%	\end{center}
%	\end{figure}
Venus's eagle wings  
signify fame.
We know she is  a housewife because 
Brant calls her sarcastically
 ``Frau Venus.''
The caricature in her hem is crude, but since it recurs
it is
probably part of
D\"urer's graphic language.

Evidently the Angel too is presented as a housewife,
not in so many words but by her
 purse, keys, and what seems to be a spindle dangling down her dress from
 her left knee.
 The {\em Madonna by the Wall} (D\"urer 1514)
also wears the keys and purse
of a housewife.
 D\"urer seems to say  that the  
 Theological Philosopher too attracts fools.
 This echoes doctrines already expressed
by  Nicholas of Cusa
\cite{CUSA}
and
Erasmus
\cite{ERASMUS}%
and 
later by
Agrippa
\cite{AGRIPPA}.

The  quaternities of  the Prophets  and of the Ghosts are also
in the theological realm.
It is not clear at this point  
whether D\"urer means quaternity to imply divinity.
First Fool provides a useful litmus indicator for this.
If the Fool  is a quadruple fool then likely
D\"urer  does not reserve
quaternity for the  divine.
So I looked for other Fools.

$\bullet$
It is then easy to see the larger face of
Second Fool looking at us and slightly to our left.
Second Fool shares curled hair  with First Fool.

$\bullet$
Third Fool is looking upward and to our left
and has a body,
a complete arm with a mitten-like hand at his side,
a leg,
and the purse of the Angel.

This set me to search for ambiguous trinities
in the Intellectual World
and  to find
Third Ghost, already mentioned.
But in fact there are four Fools and four Ghosts.

$\bullet$ 
Fourth Fool stands upright to the left of the others,
wearing a robe from head to ground,
looking almost right at us.

\subsection{Purse and Keys}

The  keys and purse of the Angel,
the only elements in the engraving
whose meaning has been left us by D\"urer,
stand for power and wealth.
But
while his Madonna and his Venus wear their
purses and keys at the waist,
the Angel drags her purse on the ground.
Likely this means that fiscal power 
profanes theology,
for D\"urer was a passionate critic of clerical corruption.

\subsection{Compasses}

Whenever the quadrivia are personified, 
Geometria gets the compasses. 
Since the Angel has the compasses, some say that
she might be Geometria.
In
a D{\"u}rer woodcut of 1504, however,
{\em The Astronomer} measures a globe 
with compasses under a full moon, 
so compasses 
also occur in the Celestial World.
D\"urer's compasses do not characterize their bearer definitively.

Some say the Angel has measured 
the stone Globe that lies before the dog. 
But the Astronomer studies 
his globe intently while the Angel looks right past this globe
into space;
as indeed she should if she is Contemplation or Theological Philosophy
and the globe is 
the Elemental World.
Lines of sight matter in the work of a projective geometer. 

To use compasses one must
control both points at once, either by holding the apex 
or both grasping both points.
The Angel  grasps only one arm of the compass
and that near the point, and she has no drawing board or table beneath the point,
so she cannot use them
as compasses.
If anything, she seems to be
sticking herself in the thigh and smiling.
Since this is absurd we must look deeper.
A few centimeters deeper will do.

\subsection{Three Prophets and a Serpent}

\label{sec:PROPHETS}
 
$\bullet$
The Angel's right knee, showing a bright triangular patch in the moonlight,
is also the top of a hood worn by
a  large subliminal man.
For reasons that will become clear 
I call him Third Prophet.
His profile is cocked at
about $45^\circ$ to the left.
He has a full black beard, an aquiline nose,
Arab attire, and
a black headband across his brow.
His face eventually leads us to his powerful shoulder,
almost a hunchback,
on which Angel rests her elbow
and compass point.
His left arm reaches toward 
the saw/sword on the ground.
The left foot of the Angel,
where it extends beyond her robe, is also his left 
hand, and her toes indicate
his
fingers.
In the other direction his shoulder  leads us to his torso, waist,
and knees. 
The Angel steps on the hilt  with 
her right foot as though to foil his effort.

$\bullet$
Now one can see what the Angel is doing with her
seemingly useless compasses.
She is sticking  Third Prophet.
His right hand,
which doubles as her spindle,
 reaches up past his face to fend off
the compass point, indicating 
discomfort.
He is shown
from head to foot.

This reminds us that  D\"urer's parents 
may have been refugees from the Islamic army of
Muhammad II,
judging by when they left Ajt\'as.
I  am reluctant to suggest that D\"urer's
Angel 
half-smiles
as she surreptitiously pricks 
Third Prophet,
but his Knight too seems oblivious to Death and the Devil
as he  crushes one of them 
between his fists.
The sneering cliffs around the Arco cathedral
are acts of concealed aggression by D\"urer himself.
At least we can cast the Angel as the Contemplative Faculty
with more confidence,
now that we understand her anomalous compasses
and her subliminal faces.

%	
%	\begin{figure}
%	\vspace{6in}
%	\begin{center}[In preperation]
%	%\includegraphics[width=6.0in]{0PROPHET1}
%	%\includegraphics[width=6.0in]{0PROPHET2}
%	%\includegraphics[width=6.0in]{0PROPHET3}
%	%\includegraphics[width=6.0in]{0PROPHET4}
%	\caption{Four Prophets. \MI\, detail. 
%	``The Draftsman",  A. D\"urer, 1528}
%	\label{fig:0PROPHET1}
%	\end{center}
%	\end{figure}

Third Prophet  is the first element of a quaternity.
To see a second,
find  
a lighter triangle in the shading of the dress
directly below the Angel's right knee,
a moonlit
 isosceles triangle next to the face of Third Prophet.

$\bullet$
Turn your head through 45$^\circ$ clockwise
from the vertical,
and see  this lighter triangle  as something like a horn.
It leads  to the dark line of a mouth in the face of a man
in a skull-cap
in a full front view.
I call him
Second Prophet.

$\bullet$
In the same view as Second Prophet is
an older and hairier subliminal face 
in partial profile looking to the left with 
a hooked nose, a sour expression,  and a skull-cap:
First Prophet.
Young Second Prophet and old First Prophet  share an eye.
First Prophet blows a shofar, a ram's horn,
as is done on Rosh Hashonah, the Jewish New Year.
As we shift our interpretation of the same view from Second Prophet to First Prophet,  
 Second Prophet's  Serpent  becomes 
 First Prophet's  nose, Second Prophet's horn becomes part of
First Prophet's horn, Second Prophet's mouth 
becomes the rest of First Prophet's horn.
 
 $\bullet$ The fourth member of this quaternity is
 connected to
 Second Prophet. 
In \MII\, it  is  a small triangular
face completed contained within that of 
First Prophet, sharing the aquiline nose of First Prophet
 but not the eyes.
 Indeed, it may be the head of an eagle.
In  \MIO\,  it is a serpent.

This quaternity in the Angel  is hidden by the bricolage technique
as well as the watermark technique.
Its elements
belong to a larger
{\em Gestalt}, the robe of the Angel,
 which leads us to give them a totally different
interpretation.
Again
the four faces can be seen only one at a time.

We already read the Angel 
as the Contemplative Faculty  of the theologians.
D\"urer seems to show by this quaternity that 
her entourage includes Jews and 
Muslims as well as Christians,
so Contemplation does not lead all her practitioners to
one absolute truth.
The sword for which the Muslim is reaching suggests 
that in this engraving D\"urer 
associated  Islam with militancy.
This is plausible. 
Hungary  had been mauled by both  Christian and  Islamic 
combatants  in the 15th century.
The first response of Muhammad II to the Crusades
was barely stopped 
by the forces of John Hunyadi,
 who became
 a Hungarian national hero.
Later the Ottoman Empire would include 
Buda and reach to Vienna for a time.
 Since the fall of Constantinople was so
epochal for Christianity,
D\"urer
might conceivably have  had
Muhammad II in mind.
But since Muhammad II was more
warrior statesman than theologian,
D\"urer would associate him 
with the Rational Faculty of the Celestial World
more than
the 
Contemplative Faculty of
the Intellectual World.

A more plausible  
original for Third Prophet
is the 
Prophet Muhammad himself.
This 
suggested  that the three Prophets of the Angel
are
 the founders of the three major 
monotheisms,
Moses, Jesus, and Muhammad,
drawn in 
chronological order from left to right.
This would be proper company for
the Contemplative Faculty
 of the theologians.
In the engraving as in the sacred texts,
the first  two of the founders  were Jews, 
the first
 lived to old age,
the second died young
and was identified with God,
and  
the third was Arab and 
a great warrior.
 The probability that all these  
agreements are chance does not seem
prohibitively large
so I follow this interpretation in what follows.

The {\em Occult Philosophy}
 gives cosmic significances
to each of the numbers from 1 through 10.
Agrippa gives by far the most meaning to the number 4:
the ancient Pythagoreans swore their oaths
on it,
there are four elements, 
and so on.
This might have influenced D\"urer to show quaternities of faces and 
the $4\times 4$ Table over other possibilities.

In  earlier works of D\"urer
with subliminal faces the faces are erect.
D\"urer varies the orientation of 
at least some of the subliminal faces in \MI\,
to make us
examine this engraving,
unlike all his others, from various angles and see
that
 the Octahedron 
changes its apparent shape.
This
illustrates 
a central point of the engraving:
Absolute Truth is not available to us.
In this D\"rer parts from Agrippa,
who finally finds Absolute Truth through his Gateway III.

\section{THE SECOND WORLD}

\subsection{Millstone} %051015

The most mysterious of the three stones in 
\MI\,
is the stone wheel
on which the Boy sits.
Panofsky and Yates call it a grindstone
and Doorly calls it a millstone. 
Its perimeter is broken, 
its face is smooth.
It would make a passable millstone but an oversize 
bone-rattling grindstone, so
surely
Doorly is right.
This is also the only stone in the engraving that could serve
Jacob first  as pillow 
and then as pillar, if he could erect it.

Since the stone Globe
 stands for the Elemental World
and the stone Octahedron for the Intellectual World,
by elimination it would seem that the Millstone 
should somehow stand for the Celestial 
World, the realm of  Reason
and  Mathematical Philosophy, or astrology, 
Gateway II. 
This representation would have to be obvious to his intended
viewers as well.
How is a Millstone like the Celestial Sphere?

Both turn slowly.
In olden times
there were few   
objects that rotated slowly
about a nearly stationary axis,
and the millstone is one.
It seems that for centuries before D\"urer,
cultures as diverse as
Babylon, Greece, Arabia, Scandinavia, and Rome 
represented the Celestial World by a millstone.
Later, Galileo  would too:
``Next, applying this reflection about the millstone to the stellar sphere, ...'' \cite{GALILEO}.

In a widespread myth,
the celestial millstone wanders off its axis into the sea,
presumably the Milky Way,
to represent the slow precession of the 
equinoxes from one house of the zodiac
to another \cite{SANTILLANA}.
Santillana and von Dechend called it
 Hamlet's Millstone
because it is mentioned  in passing
 in the legend of Ambleth, the original Hamlet,
as retold by Saxo Grammaticus  
in the late 13th century.\cite{SAXO}.
Saxo was first printed 
in Paris in the very year 1514 of \MI,
indicating that the trope of the celestial millstone  was still
alive in D\"urer's day.

According to the myth, in the Golden Age the 
millstone ground out gold.
That is, 
in the good old days
the stars were astrologically propitious.
In later, lesser times it ground out salt,
making the ocean undrinkable.
In our decadent age it produces sand
and the terrible maelstrom,
whose very name 
invokes the myth.
The fault is in our stars, not in ourselves.

Saxo's Amleth and
Shakespeare's Dane share with
D\"urer's Boy the themes of
melancholy,
parental Ghosts, 
and dynastic overthrow, besides the millstone.
This suggests that  D\"urer and Shakespeare
both transmuted 
personal grief into the Agrippan creative melancholy
that inspires  great works,
the one mourning his
mother Barbara,
the other his son Hamnet,
and that both infused their ensuing creation with
Saxo's tale of familial
bereavement,
whether independently or not
I cannot say
\cite
{GREENBLATT}.

The Faculties
are out of their proper spheres,
but since a revolution is going on in this engraving, 
 some rearrangement in the seating  is understandable.
 The seated Angel, the Contemplative Faculty, is
nominally of the Intellectual World III, the
highest, but
here she is grounded and  lowest.
D\"urer has demoted Theological Philosophy.

The Boy, as the Imaginative Faculty, is
properly of the Elemental World 1,
but here he perches on the Celestial World II.
D\"urer has promoted  Natural Philosophy
above Theological.
The natural philosopher
observes matter in space and time,
traditionally the lowest World,
but thereby approaches God.
The impending scientific revolution  is in the seating plan.

The Boy watches passively and draws what he sees.
We know that D\"urer himself did experiments, since he invented 
drawing aids and etching,
but he has not made them part of his philosophy or his Boy.
The evangelist of experiment, Francis Bacon, has yet to be born.

{\em Pace} Yates,  the equity  between 
the Angel and the Boy, therefore,
is not the Hermetic equality
between the worlds above and below.
Hermetic doctrine relates  the Elemental World to the Celestial,
and becomes irrelevant when the celestial millstone is taken out of service.
``As above, so below'' may well have been 
Agrippa's credo
but it is not D\"urer's. 
Like Leonardo, D\"urer  
caricatured the idea that events in the sky foretold events on Earth.
He prepared and published  printed maps of the earth and the celestial sphere as  globes.
They are realistic maps of the stars above and the planet below  that
do not indicate any similarity or correspondence between them.
The unity of Elemental and Celestial Worlds that D\"urer
symbolizes by pulling out the Millstone does not harken back to
Hermes Trismegistus but forward to Newton.

\subsection{Dog}

The Dog was especially obscure to me from the start.
\MI, {\em The Knight}\/, and {\em Jerome}
each  have a dog and an hour glass.
The hour-glasses remind us that life is brief.
What do the dogs mean?

The Dog of \MI\, dozes rather too close to the straight line
of Comet, Octahedron, and Globe for accident,
just where the Celestial World would lie
in the old cosmology.
Someone leaned the battered old Millstone 
up against the House
and the Dog 
lay down where it had been.
The talents of 
the Celestial  World were Mathematical Philosophers,
astrologers, statesmen,
kings;
the connection of the Dog to the Celestial World
was even less obvious to me  than that of the Millstone.
To us,  dogs represent love and fidelity, 
to Agrippa, they represent love or flattery,
but to D\"urer, somehow, they represent
the Rational Faculty of Agrippa's World II.
We may infer this
by elimination and guesswork,  but
we have to understand how this  could have been
understood by D\"urer's viewership.

To read this Dog I first looked at D\"urer's other dogs.
He
had drawn at least two in the year before
that are still extant,
in the {\em Triumphal Arch}
and the {\em Hieroglyphica}\/.
The dog drawn by D\"urer for the {\em Hieroglyphica} 
resembles the one in 
\MI\,
in its  gaunt belly
and its unusual head, which some take for a sheep's.

In his {\em Triumphal Arch}
D\"urer represented
 the Emperor Maximilian I as a dog with a stole,
as the {\em Hieroglyphica} required.
In the {\em Hieroglyphica} 
we learn that the Egyptians represent a prophet 
by a dog,
``because the dog looks intently beyond all other beasts
upon the images of the gods, like a prophet.''
\cite{HORAPOLLO}.
The Dog is actually a prophet,
worker of the Celestial Sphere,
an astrologer.

Reading the language of {\em Hieroglyphica} 
is trickier than writing it.
Besides a prophet, 
Horapollo's naked dog can also represent an embalmer, the spleen, odour, laughter,
sneezing, and rule.
Apparently one goes by context. 
I think the Dog means the prophet this time.
Others have opted for the spleen,
influenced by humoral psychology,
but this does not  fit as well.

The main people associated with the Celestial World
and with inspired Rational Faculty, according to 
Ficino, are
the prophets and the kings.
The {\em Hieroglyphica}
represents both  by dogs, naked or stoled.
So D\"urer obediently represented them accordingly on 
the {\em Triumphal Arch} and 
in \MI.

The dog 
as the Rational Faculty of the Celestial World fits all
three Master Engravings
rather well.
The dog of {\em Jerome} can sleep 
soundly because Jerome's melancholia inspires his third
Faculty, theology,
not his second, prophesy.
The Boy's Dog rests with half-closed eyes because his World no longer exists.
The dog of the {\em Knight} is wide awake and
runs beside the Knight
because the Knight concerns himself with affairs of state,
with inspired Faculty II.
The Knight may be
Erasmus, who believed in astrology \cite{BROSSEDER}.
According to traditional Catholic apologetics 
St. Jerome did not.
This increases  confidence in the interpretation.

\subsection{Wave
} 
\label{sec:WAVE}
%060222

My
 Bible-searches for ladders, heavenly lights, 
and sky bows all bore fruit
so following my hermeneutic rule 
I looked for  D\"urer's Millstone first in the Bible.
Of
12 millstones in the  Christian Bible,
all but one are metaphors for  livelihood or industry and 
refer to no other element of the engraving, so I put them aside.
The eleventh is  
in {\em Revelation}\/:
\begin{quote}
And a mighty angel took up a stone like a great millstone, and cast it into the sea, saying, 
Thus with violence shall that great city Babylon be thrown down, and shall be found no more at all.  ({\em Revelation} 18:21, King James)
\end{quote}
Four points of the engraving occur in one sentence: 
the mighty angel, the millstone,  the sea,
and even a kind of melancholia,
for {\em Revelation} imbeds
this account
in a score of verses lamenting
the destruction to come. 
The millstone of {\em Revelation}
is therefore a reasonable
candidate  for D\"urer's millstone, 
and the only one in the Bible.

But 
the Biblical millstone is as puzzling as D\"urer's.
A more cosmic image 
than
sinking a millstone --- say, sinking a continent --- 
would show how great a force would destroy Babylon.
A more domestic image --- say, stepping on a wine-glass ---
would show how easy it would be for God to destroy Babylon.
But a millstone is neither here nor there.
A mere man can sink a millstone in the sea, 
with some effort,
and it does  the millstone no harm.
The angel  of {\em Revelation} seems merely peeved with Babylon.

If on the other hand the millstone of Revelation  too
were the Celestial World,
throwing it into the sea would be
the same cosmic disaster
as Ambleth's millstone rolling into the sea,
the wobble in the celestial world that
ends one world Age
and begins another  \cite{SANTILLANA},
a simile fit for an empire.

Any  reference to the celestial millstone in the Bible 
would well be veiled,
because it is a vestige of the very
Babylonian  astrology that
Abraham fled.
The separation of the waters below from the waters above
 by the celestial sphere
in {\em Genesis} might be another biblical trace of the
 three-world cosmology.
I tentatively adopted the celestial interpretation 
for both D\"urer's millstone and 
the one in {\em Revelations}\/.

Then Babylon
and its downfall might occur in the engraving too.
The engraving shows a city by the sea
with ships in port.
Over the left-most pan of the Scales,
through a small triangle framed by Ladder, 
 House, and  one pan, 
 a closely spaced  system of fine parallel wavy lines 
flows from behind the House leftward into
a great Wave that
    looms high over the city, 
about to crash down on it and wipe it out. 
It has
already been noted \cite{HECKSCHER}
 and cited  \cite{SCHUSTER}.

%	\begin{figure}
%	%\vspace{6in}
%	\begin{center}[In preperation]
%	%\includegraphics[width=4.0in]{0WAVE}
%	\caption{The Wave. \MI\, detail.}
%	\label{fig:0WAVE}
%	\end{center}
%	\end{figure}
It is easy to overlook  the Wave
 because
it is ambiguous.
It can also be
merely  the shoreline beyond the city,
not looming over the city at all.
Then the  lines are merely geological strata, not water.
We cannot tell whether they are near the city or far
from one perspective view.
The ladder  hides 
connections that would force one interpretation or the other.
If the shore had a smooth C-shaped curve 
we would tend to see it as horizontal, but
D\"urer has made the visible part of the shoreline
oddly straight and vertical on the canvas,
evoking
our propensity to recognize vertical lines.
 He has also shaded the Wave darker than the remote hills,
 to allow us to separate them.
He
allows us to 
see this line either as a vertical wave or a horizontal shoreline.

 The Wave
 brought me
at last 
to the realization
that this recurrent perspectival
ambiguity was not a weakness of D\"urer's art
but 
the very point of the engraving,
driven home again and again
until even I could not miss it.

\subsection{Destroyers}

 In  \MIO\, the ambiguous vertical shoreline-or-wave-front 
 framed between
 the fourth and fifth rungs of the ladder
 can also be seen as a face of a man or woman.
Then it is not hard to find a full quaternity 
of Destroyers of Babylon.
They are present but completely reworked in \MII\,
so there can be little doubt that they are intentional.
The most evident change is in the lip line
of the leftmost First Destroyer, whose profile is
 the shoreline or wavefront.
With  a single slant stroke of the burin
D\"urer  changed
a light smile in \MIO\, to
a dour frown in \MII,
as better befits a divine Destroyer.

Now the engraving and the  verse of {\em Revelation} share 
Millstone, Angel, sea, Babylon,  destruction, and 
melancholy.
This reinforces the connection we sketched in lightly between 
D\"urer's millstone and Hamlet's.
We provisionally assume that the millstones of {\em Revelation}, 
\MI, 
and Ambleth,  centuries and much of 
a continent apart,
all represent the Celestial World,
and that D\"urer intended  the
engraving to refer to {\em Revelation}\/, the end of his Bible,
as well as {\em Genesis}, the beginning.

The Biblical verse would have referred to the end of the 
Age of Aries and the beginning of Pisces.
If D\"urer meant the reference seriously,
he must have referred to the end of Pisces and the beginning of Aquarius.
Since D\"urer had already executed the hierarchy of the Church {\em in absentio} in his {\em Apocalypse}\/,
he might have meant the same by
the destruction in \MI. 
Then the picture is in the future tense, as {\em Jerome}
is in the past and the {\em Knight} in the present.
This tells us how to hang the triptych.
The {\em Knight} goes in the center and \MI\,  on the left.

\section{THE FIRST WORLD}

\subsection{Limen caelo} %05080

The name for the engravings  is taken from
the motto 
that appears
on a banderole. 
Actually, the motto is   
``MELENCOLIA\S I",
with the symbol {\S}{}, a fancy S,
in a lighter style than the rest.

{\it Melancholia} is the transliterated  Greek 
for melancholy and black bile.
One Latin form is {\em melancolicus}\/;
The German is {\it Melancholie}\/. 
The Italian was {\em Malinconia}\/.
D\"urer himself uses the etymologically reasonable
form {\em Melancolia} 
in a title-page woodcut of 1503.
In no language 
is the word {\em melencolia},
a spelling  
odd even for D\"urer.
The spelling of the time was so variable
that we cannot be sure, but this spelling
may be imposed on the overt message by 
a covert one that D\"urer wished both to record and conceal
in an anagram.
The encodings we have already found,
 the misspelling, and the cryptic appended symbols {\S}I suggest 
 this possibility.

Agrippa could well have
put anagrams, gematria, and  magic squares
into D\"urer's mind.
They are all  in his {\em Occult Philosophy}\/.
They
 are  also among the 
techniques used to squeeze new meanings out of old texts  in the
{\em Baraita of Thirty-Two Rules}
of Rabbi Eliezer ben Jose the Galilean,
who flourished in the second half of the second century C.E.
and was cited importantly in later years.
Such lore became more accessible 
to  Florentine Neo-Platonists like
Agrippa,
an active Hebraist,
when  Jews expelled from Spain in 1492
settled in Florence.
 
In any case,
by the 16th century 
letter-permutation was
a standard way to protect intellectual property. 
The Royal Society and the
system of scientific archives were still over a century in the future. 
Writers of the time who solved an important problem 
could not guard their intellectual property  
by publishing or patenting.
Some 
protected  their priority  by permuting the letters of their  solution
and publishing the result together with the 
problem. 
Earlier Roger Bacon scrambled the formula for 
gunpowder in this way in order to prevent the proliferation of this terrible weapon. 
Later Galileo would scramble his discoveries of the phases of Venus and the rings of Saturn.
Mathematicians scrambled their theorems.
The resulting
meaningless jumble of letters declared on its face that it was a cipher. 
Sometimes it was misconstrued by a false rearrangement.

In the next degree of concealment,
the
scrambled letters themselves spell out  a cover message,
the anagram.
This can hide not only a message but even 
the existence of that message.
 \MI\,   does this well. 
Only the cover message has been read in recent centuries as far as I know.
I unscrambled the anagram as soon as I began to work  at deciphering the engraving, 
as follows. 

Since Panofsky declared the engraving 
autobiographical,
I tentatively supposed that the motto 
represented D\"urer himself.
To 
see how he might describe himself graphically
 I
 went to  his coat-of-arms (1490,  1523).
 
%	 \begin{figure}
%	\vspace{1in}
%	\begin{center}[In preperation]
%	%[FIGURE ABOUT HERE: 2 INCH WIDE. ]
%	\caption{D\"urer's coat-of-arms.}
%	\label{fig:ARMS}
%	\end{center}
%	\end{figure}

The 1523 version centers
 his famous monogram and the year
above his coat-of-arms.
Next below them come clich\'es 
one finds in any dictionary of heraldry\cite%
{WADE}:
a blackamoor  for heroic action 
by an ancestor in the Crusades;
eagle wings for fame and glory;
a closed helmet in full profile for the estate of
esquire, the lowest.
A shield below these clich\'es bears  an image specific to the family.

$\bullet$
  The coat-of-arms shows a  gateway with open doors.
This is self-evidently
an  ideograph for both ``D\"urer''  and ``Ajt\'os,'' 
 the family name and origin.
 
This sent me looking for other ideographs.
One is near at hand in the coat-of-arms itself:

$\bullet$
The gateway stands on a cloud,
more prominent in the 1490 version than the 1523.
The gateway is a gateway
 in the sky.
 
I read this as follows.
Coats-of-arms usually display the pride of the family,
some 
great personal accomplishment. 
To proclaim his greatest talent,
engraving,
D{\"u}rer could simply 
have drawn a burin on the shield,
or  the word ``burin" itself.
But
the Latin for burin  is
{\em caelum}\/.
This is also
 the ordinary word 
for ``the sky, " ``heaven," and ``the heavens." 
It has the same root as our ``celestial." 
{\em Caelo} means both the noun {\em (in) Heaven} and the verb 
{\em (I) engrave}\/.
The root seems to mean both swelling and hole.

$\bullet$
The D\"urer coat-of-arms
is both an ideogram  for``Gateway in Heaven" 
and  a rebus for
``I engrave the gateway."

``Gateway in Heaven" was   a well-known
metaphor for the Roman Church itself. 
A person of the time might
take the D\"urer coat-of-arms not 
as a sacrilegious display of ego but
as a pious reaffirmation 
that we are made in the image of God. 
But it can  also  serve as
a banner of Humanism:
the Gateway to Heaven is open to us all.

Gates of  Heaven  open
again and again in \MI\, itself.
The  D{\"u}rer monogram 
AD, exhibited  
at the top center of the 1523 coat-of-arms and at the bottom 
right corner of \MI, is one of them.
It seems at first to be a pun on  {\em Anno Domini}\/.
Now one can see that
it is also an ideogram.
The flat-topped 
A is a gateway as well as a 
letter. 
The legs of the A are the uprights of the gateway.
The lintel sits on them and connects them.
A stiffener just beneath the lintel is
the crossing of the A,  rather  high
 for an A but proper for a gateway. 
 The D between the uprights of the A
echoes the Doors between the 
uprights of the Gateway in the D\"urer coat-of-arms.
In  his 
coat-of-arms
he 
symbolized his holy art
by  the punning {\em caelo} rebus.

As soon as I found this pun I checked to see whether
CAELO fits
into MELENCOLIA, as it does.
The leftover letters then 
spell out LIMEN, 
commonly meaning
gateway,  a near-synonym for ``D\"urer,''
and  good description of the gateway in the D\"urer 
coat-of-arms.
It also means
gate, doorway, threshold, lintel, 
walls, house, home, boundary path, 
and limit, according to context.

$\bullet$ Unscrambled, 
the first word of the motto becomes
{\em LIMEN CAELO}\/,
``Gateway to Heaven'',
a description of the D\"urer coat-of-arms.

Panofsky was right, and
we have another signature.
Its unscrambling took no longer than it takes to describe.
This made me confident that I was
conversing with D\"urer, not with myself.
The hidden motto  combines the family name-rebus,
D\"urer's art,
and the D\"urer coat-of-arms itself.
It
supports
the interpretation that D\"urer
 intended the clouds in the coat-of-arms as a 
symbol of Heaven,
and that he did not  demonize Agrippa.
The hidden phrase also
applies to the dim archway in the heavens
that frames the motto,
and to other
elements that
we take up  later.
These multiple points of contact
between the anagram and the engraving
suggested that I had read it correctly.

The overt message and image, both dark,
cover a positive one.
``Limen caelo'' is as bright in spirit as
 ``MELENCOLIA'' is depressive.
And yet the picture is dark.
 It seems to offer
  a  gateway to Heaven, but
  we must still figure out what that is.

There remains the  ``{I}"  of the banderole.
Some suggest
that it refers to one third of 
the medieval tripartite universe, and 
expect  that   `` MELENCOLIA II`` and   III  were engraved and lost,
or were at least planned.
 \MI\,  is indeed one of three
but they all survived famously well.
{\em Jerome} and \MI\, clearly
exemplify Faculties III and I;
Agrippa's Book III lies open before Jerome
as his Book I lies open before the Boy.
Likely
the {\em Knight} represents Faculty II
as a statesman would.
We see why one engraving bore its own caption
 and not the 
other two: 
Gateway I is the one that D\"urer wishes
to announce.
And this explains the
 ``I''. 
 
$\bullet$  The full Motto ``MELENCOLIA{\S}I''   is an anagram of 
{\em LIMEN  {S}I} CAELO\/, meaning both ``Gateway One to Heaven"
and "I engrave the Gateway".

For I require that D\"urer  makes the same pun on{\em caelo}
in the Motto
that he does  in his coat-of-arms.
 
 Another usage of ``I"  would also have been natural for a
mathematician like D\"urer.
Even in 1514, an
``{\em x 
primus}"  in a mathematical writing, including D\"urer's,
 referred back to an earlier $x$
 and did not imply a later {\em x secundus} or {\em x tertius}. 
 This is true  D\"urer's own mathematics too.
Then the Motto would mean, however,
 that D\"urer was superceding the
old Gateway to Heaven with a new one,
thus  claiming innovation.
This would have provided the capital crime we seek but
I abandoned this interpretation
 because his writings indicate
that  he  did not
consider himself an  innovator but
a renovator stripping away the innovations
made by the  Church.

We do not need to postulate lost engravings
of the three  Worlds  to explain the  Motto.
\MI\, itself shows all three  Faculties, 
Worlds, and  Gateways to Heaven.
The Boy 
represents the  Imaginative Faculty I choosing Agrippa's
Gateway  1, the Natural Philosophy
that evolved into Physics.
For D\"urer it 
meant making precise and beautiful records of nature
and knowing and obeying the laws of perspective and structure. 
Gateway II is
 the way of Mathematical Philosophy  and the Prophet,
 represented by the Dog.
Gateway III, the way of Theological Philosophy,  is
the way of the Angel.

\subsection{Ladder} %050803

Any artist who put an endless ladder next to an angel, a house, 
and an 
up-turned stone in 1514 could be sure that the
 educated viewer of the time 
would see 
Jacob's ladder, angel, house, and stone of {\em Genesis} 28.
The  upper end of the Ladder is unseen, so it may be in heaven,
but is surely not the ladder of D\"urer's crucifixion scenes.
Ladders as gates of heaven occurred 
commonly in earlier art, including D{\"u}rer's own. 
The most relevant verses are
\begin{quote}
 And he dreamed, and behold a ladder set up on the earth, and the top of it reached to heaven: and behold the angels of God ascending and descending on it. 
 {{\em King James Bible,  Genesis} 28:12}

And he was afraid, and said, How dreadful is this place! 
this is none other but the house of God, and this is the gate of heaven.
 And Jacob rose up early in the morning, 
and took the stone that he had put for his pillows, 
and set it up for a pillar, and poured oil upon the top of it.%
 {{\em King James Bible,  Genesis} 28:17-18}
Ê  \end{quote}
So
the ladder is the Gate of Heaven  of 
 {\em Genesis} and also of the coat-of-arms; it is another signature.
{\em Beth El}, the House of God, is also the legendary site of the first house of worship 
of the God of Israel. 
Early on, therefore,  {\em porta caeli}, the Gate of Heaven in the Vulgate, became a metaphor for the Catholic Church itself.

\subsection{House} %051015

The house too is semantically polymorphic. 
From up close we see two blank walls 
supporting a miscellany of instruments. 
The side wall does not seem much wider than the Boy 
leaning against it. 
It seems about the right thickness for an out-house or a chimney, not a house.
Its front wall extends out of the scene to the right and holds a Bell, 
a square array of numbers,, 
an Hour-glass, and a Sun-dial. 
Its left edge bisects the picture to the millimeter.
Its side wall holds a chemical balance or Scales.
The Bell-rope trails off the edge to our right.
Step back several meters or squint
and the picture changes.

$\bullet$
Square and Bell together become
a lattice window, the Hour-glass 
a bay window, the Scales  a side window, 
and the out-house  a full-sized house showing us 
three of its windows.
The Angel blesses all three windows with her wings.

Assembling familiar objects from unrelated parts in this way, 
for example, faces 
from fruits, or demons from vermin, 
was an optical illusion often practiced by artists of the era.
Here there is allusion 
as well as illusion. 
If the house is what the sanctified instruments on its walls open into,
 it is the 
humanly perceptible universe, the  physical cosmos,
Agrippa's Book I.
The Ladder and the Angel have already 
told us that this is also the House of God, 
and the instrumental windows tell us that measurement and 
arithmetic look into it. 
Leonardo wrote that
knowledge comes from experience,
and D\"urer draws that scientific experience leads to God.
That the Elemental World is a House of God seems to be the teaching of 
Agrippa about Gateway I, and a Humanist idea.
This  juxtaposition of science and religion is surely not Church teaching of the time,
which provides one and only one gateway to absolute truth,
but expresses D\"urer's known Humanism.

\subsection{Magic Square} %050803
\label{sec:SQUARE}

The matrix of  numbers  set into the masonry wall
in  \MI\,
 like a window lattice 
is a magic square, meaning that 
 it is filled with consecutive numbers starting from 1 
and 
every row, every column,  
and both main diagonals add up to the same number.
In a $4\times 4$ magic square 
that number must be one quarter the sum of the integers from 1 to 16,
or  34.
This  is  a gnomon magic square.
Gnomon means that besides being magic,
 its four quadrants, its four corners,  
and its central tetrad
add up to the same number.
Usually a gnomon is a pointer or indicator, especially of a sundial, but
a carpenter's square was also called a gnomon, 
perhaps because it looks like a crude picture of
a pointing hand.
This  is called a  gnomon magic square, I guess,  
because removing a quadrant
leaves a figure that resembles  a carpenter's square. 
In addition, the sum of any pair of numbers symmetric about the center of this square
is  17.

In his {\em Occult Philosophy}
 Agrippa assigns a magic square --- 
 then called simply a table ---
to each of  
the seven ``planets" then known,
in the ancient
order of  Saturn, Jupiter, Mars, Sun, Venus, Mercury, and Moon.
This is
the order of their apparent periods about the Earth;
in the geocentric system  the Sun acquires
the heliocentric period of the Earth. 
He gave each table in both Arabic and Hebrew numerals.
These magic squares have ninth century Arab 
sources \cite{PANOFSKY}
but 
may be older.

Agrippa omitted the  1-by-1 magic square $[1]$  
--- reserving it for God? ---
and
there is no 2-by-2 magic square, so  the Table of  Saturn  
is  3-by-3, the Table of
Jupiter is
4-by-4
and so on to the 9-by-9 Table of the Moon.
The number of Jupiter is then $1+2+\dots+16=136$,
as the number of the Sun is  the  $1+2+\dots+36=666$ of biblical fame,
raising the possibility that the nine planetary tables are older then 
the biblical passage about the Number of the Beast.
Agrippa warned that unshielded Saturn  caused
 acute melancholia,
a clinical mental illness, 
and prescribed wearing
Jupiter's Table as 
shield.
Agrippa proved  the virtue of Jupiter's Table by
gematria,
a Hebrew numerology based on the fact that any letter of 
the Hebrew alphabet is also a number
and that therefore every Hebrew word has a number, the sum of its letters.
He used gematria to find
reassuring Hebrew words in Jupiter's Table.

As a child I wondered why such a square was called magic.
The {\em Occult Philosophy} answers this question at least.
They were used as magical talismans.
For example, 
Agrippa prescribed wearing the Jupiter Table to counteract
the melancholy influence of Saturn.
Some say D\"urer borrowed Jupiter's Table from 
Agrippa \cite{PANOFSKY, YATES1979}.
Obviously he did not:

$\bullet$
The tables of Jupiter
and D\"urer 
have no row or column in common.

He performed non-trivial
permutations on Agrippa's Table of Jupiter
--- not mere reflections and rotations of the whole table --- 
that left it both magic and gnomon in the arithmetical sense.
Namely, he interchanged 
the first row with the last,
and also the first column with the last.
This
changed
the bottom line to 4 15 14 1.
 This  is
the date 1514 of the engraving, flanked 
by D\"urer's initials in the $A=1, \, B=2, \dots$ code, the Latin gematria.
It echoes the by-line below it.
It tells us something about D\"urer's mind that he 
was able to see this line as a possibility in the Jupiter Table.

\subsection{\MII}

There are at least two engravings by D\"urer in circulation 
called \MI, though
I have never seen this pointed out.
The two
differ most
conspicuously in the D\"urer  Table.
In  \MIO\, displayed on the internet (for example) 
by the Fine Arts Museums of San Francisco,
the numeral 9 in the Magic Square
starts out much  as usual,
though it ends with a wriggle in its tail that makes it more like a 
question mark than a numeral.
In \MII\,  shown on the internet by the British Museum,  the 9 is reversed,
more  an S  than a 9.
The interpretation suggests
that D\"urer made
 \MII\  after
\MIO;  hence my names for them.

There are numerous other differences as well.
In \MIO\, the left side of the Angel's lips (our right)
ends in an upturn, either of the lip-line or a sleeve-line.
In \MII, this part of the engraving is simply blackened.
\MII\, also reworks some of the robe of the Angel 
and the Wave.
I return to these changes and point out others
in due course, for they reveal
D\"urer's concerns.

 In Section
    \ref{sec:FLOURISH}
I propose one 
  interpretation in due course
   that works for  all eight variant forms in the Table
  and the Flourish in the Motto.
  
%	  \begin{figure}
%	\vspace{1in}
%	\begin{center}[In preperation]
%	%\includegraphics[width=2.0in]{0TABLE1}
%	%\includegraphics[width=2.0in]{0TABLE2}
%	%\includegraphics[width=2.0in]{0TABLE3}
%	%\includegraphics[width=2.0in]{0TABLE4}
%	\caption{Jupiter's Table (leftmost)
%	and  D\"urer's,  variants A and B}
%	\label{fig:0TABLES}
%	\end{center}
%	\end{figure}
 
\subsection{Door}

Then there is the curious matter of the door to the House:
there is none.
The House that we see as we step back has several windows
but nowhere do we see a 
door. 
When Mr. Door draws the House of God
without a door
he is likely making a  
statement.

It seems to repeat the major statement of the piece.
If the house is indeed the absolute reality of
theological philosophy,
and it has windows but no doors, 
then we can look but not enter.
If  it can be entered  by the ladder,
that is for the angels.
The inaccessibility of absolute truth
was a common idea by D{\"u}rer's time,  
taught for example
by  Nicholas of Cusa and Erasmus, already cited,
so
D\"urer could expect some of his viewers to
make out this message at first sight.
Before D\"urer did this engraving he had 
already written that the human mind 
cannot know absolute beauty.
We must not assume 
that he separated truth and 
beauty
 as cleanly as some claim to do today;
he  expected a mathematical 
theory of both.
In this picture he seems to 
be saying the same thing
in several ways.

\subsection{Boy} %051015

$\bullet$
The Boy's writing  instrument has a crossbar at its top, and
so is not a stylus 
or a piece of chalk --- as some have said ---
but  a 
graver, the celestial burin again.  
His block is therefore not a slate but a copper sheet or a block of wood.
He is engraving, not writing.

The Boy's eyes are disturbing from
close up.  I cannot see them as both  looking
in the same direction.
His left eye could be looking down at the Globe, while
in  \MIO\, a ring of white in his right eye
gives the strong impression that the Boy looks
directly as us with that eye wide open.
 \MII\, changes the ring of white into
a more naturalistic crescent of white,
but the Boy is still looking at us with his right eye and down with his left.
%	\begin{figure}
%	\begin{center}[In preperation]
%	%\includegraphics[width=4.0in]{0BOY}
%	\vspace{1in}
%	\caption{The Boy.  
%	\MI,  A. D\"urer, 1514, detail}
%	\label{fig:0BOY}
%	\end{center}
%	\end{figure}
D\"urer himself had amblyopia.
The Boy then carries  two distinctive
attributes of D\"urer, his burin
 and his amblyopia.
 D\"urer drew his mother's portrait with divergent amblyopia,
like  Third Ghost,
but the Boy exhibits a convergent amblyopia.
Perhaps his condition was a wandering eye.

Panofsky was righter than he realized when he said that
D{\"u}rer meant this work to show many aspects of himself.
The Boy is no generic starving artist scribbling 
meaninglessly on a slate, as Panofsky suggested, but is as Yates said,
  D\"urer himself, at work with his heavenly Burin,
his strange eyesight,  and his art of perspective
in the creative frenzy
that melancholia can inspire in the divinely gifted artist.

He
is the central figure in several senses.
The geometric central axis of the engraving is the
vertical defined by the leftmost edge of the house.
It passes almost exactly through the nearest eye of the Boy,
a common way to indicate
centrality in Renaissance art.
The line of the comet suitably extended
also strikes the Boy's left eye.
 
 Yates suggested that the Boy 
 D\"urer is engraving
\MI\,
  itself in a
 self-referential way. 
 The Boy's copper plate is too small for that
 suggestion
 to be taken literally.
 
The Burin identifies the Boy as D\"urer.
In the tripartite psychology of the times,
the artist's special Faculty is the Imaginative one,
that of the Natural Philosopher
so the Boy  represents Natural Philosophy and the Imaginative Faculty too,
as the Angel represents Theological Philosophy.
Then he should be looking at the Globe,
the world of the four elements in space and time.

\subsection{Four Friends}

$\bullet$
if we cock 
our heads about 30$^{\circ}$ to our left,
we can see or imagine an erect face
peering from behind the Boy, 
in the part of the blouse
billowing out between two straps.
 I call him
First Friend.
%	\begin{figure}
%	\begin{center}[In preperation]
%	%\includegraphics[width=4.0in]{0FRIENDS}
%	\vspace{1in}
%	\caption{Four Friends.  
%	`\MII. A. D\"urer, 1514, detail}
%	\label{fig:0FRIENDS}
%	\end{center}
%	\end{figure}
His head is bounded on our right by the curve of
the Boy's gown between the straps
and is looking downward towards the viewer's
left.
The Boy's gown, where it passes over the Millstone,
forms two near right-angles
that look uncannily like a small book, and First Friend seems to have 
a hand on it.
That and the robe trailing behind him make First Friend look clerical or at least scholarly.

First Friend too is part of a possible quaternity
in  \MIO\, that D\"urer substantially reworks in \MII.

$\bullet$ To see the profiles of Second Friend  we should stand the engraving on its left edge.
The fringe of the Boy's garment then becomes the profile of
Second Friend.

$\bullet$  Third Friend is the Boy's left upper arm.

$\bullet$ In \MIO\, Fourth Friend is the Boy's right lower leg. 

$\bullet$ In \MII Fourth Friend is a smaller better-formed head set into
the Boy's right knee, touching the Boy's block.

\subsection{Scales}
On the side wall of the house hang a pair of chemist's scales.
Extended, their lines pass through the vanishing point,
so the Scales are balanced. 
Perspective 
causes the line of the pans on the paper to slant slightly upward toward the vanishing
point, and the line of the suspension slightly downward.
One pan brushes the Boy, the other the Angel.
I think we can say that D\"urer has established a balance between
Natural Philosophy and Theological Philosophy,
despite Agrippa's flight into fideism.
The line of the eyes of Boy and Angel also goes through the vanishing point.
They have exactly the same altitude,
that is, holiness.
The two remaining philosophies are equally holy Gateways to Heaven.

\subsection{Globe} %050513
Chance governs the Elemental World,  mathematical 
law governs the Celestial World, 
divine law  governs the Intellectual World.
D{\"u}rer's  goddesses 
Fortuna and Nemesis
both stand on globes, liable to roll 
in any direction at any impulse, 
because Luck and Fate are unpredictable
\cite{PANOFSKY}.
Therefore D\"urer chooses a Globe 
to  represent the Elemental World. 

It is 
indeed the lowest of the three stone objects in this picture, 
as it should be if
they represent the Three Worlds.
D{\"u}rer, like Nicholas 
of Cusa and Leonardo, 
has replaced the naive tri-spherical geocentric Neo-Platonist cosmology by 
a tri-partite philosophy with three Faculties and Worlds that are
no longer geocentric, concentric, or even places at all.
And he has retired the Celestial World.

\subsection{Moonbow}

I have never seen a moonbow but again  NASA  
provides the picture and the Bible provides the caption.%
\cite{METEOROS} is an on-line gallery of portrayals 
of heavenly phenomena 
like moonbows and moon halos, including several from 
D\"urer's time. 
The only bow in the Bible sky is the ``bow in the clouds''  
that Noah saw after the flood,
a supreme example of divine revelation,
direct communication with divinity.
{\em Genesis} does not say whether Noah saw his bow 
by  day or by night.
D\"urer had no choice
once he set the scene in the dark of night. 
 In {\em Genesis}\/, the  light in the heavens  and the bow in the clouds 
represent two of the great gifts of God,
so  the picture should be  joyous,  and yet it is dark
and ambiguous
again.

The viewer of any moonbow
or rainbow lies on 
the normal to the  plane of the bow that
passes through the  center of the bow.
D\"urer's watching eye 
should therefore lie directly before the horizon point
beneath
the highest point of the moonbow.
This is not exactly where he is located by
the vanishing point of the building lines,
which is under the eye of the Chimera,
but it is only about a centimeter off.

When we look directly at a bow in the clouds,
the light source must lie directly
behind us, and
very likely D\"urer knew this from 
personal experience.
The artist is standing before the Chimera,
a bit to the left of the center of the Bow.
Therefore to make such a bow the moon should be
 behind him and slightly to
his left.
But according to the shadows on the House 
and the face of the Angel
the moon is behind the artist 
and well
to his right.
This optical inconsistency 
seems to have no meaning, 
but we need not try to believe 
that D\"urer made a mistake in his depiction of
nature.
Had he followed 
natural law
the bow would have been far out of the picture
 to our left.
He wanted to show 
shadow on the Angel's profile
to at least suggest melancholia,
and God's promise in the sky
to express faith and inspiration,
a physically difficult or impossible combination in any one view.
I suspect that he deliberately 
sacrificed optical truth for a greater artistic one,
knowing that few 
would notice or care.

\subsection{Comet} %051015

Though it lacks the curved tail
of the most traditional renderings
the bright Comet in the sky
is a comet, not a nova.
The NASA online collection of astronomical photographs
shows comets quite like it.
A comet is straight when it heads for the sun
and sunlight pushes its tail  out behind it.

The great comet 1471Y1 was first seen on Christmas Day in 
D{\"u}rer's birth year,
and D\"urer wrote of seeing a comet himself 
in 1503
\cite{CONWAY, KHULUSI}.
The  physical natures of meteors and comets
were not yet known in 1514.
Leonardo believed, and Galileo would
still believe in the next century,
and our very words still reflect their belief,
that meteors were meteorological,
weather of the high air,
and comets too.

Since the Bible has already explained other elements of the picture,
let us search it for this one too.
The answer is swift and unique.
The terms ``comet,''  ``shooting star,'' and ``falling star'' do not appear. 
There are three cases of a star that falls in {\em Revelation},
but they do not fit the picture especially.
The  only blinding light 
in the biblical sky is the original light of the
divine creation.
Then the starless sky  would recall the time in the Bible story
between the creations of light  and  the stars.
The Comet would then represent divine creation and 
divine revelation.

The {\em Hieroglyphica} has no comet, but it has a star,
another glyph for God.
I see no way to decide whether D\"urer 
had {\em Revelations} in mind or the {\em Hieroglyphica} 
when he drew the Comet/Star, but 
the two interpretations are consistent, even mutually supportive.
The comet tail points quite accurately at the Boy's head.
This suggests that God is singling him out for illumination,
above the Angel,  Dog, or  Chimera.

  The Creation is a  joyous occasion:  
  Happy birthday,  universe!
The news is good but the tone is somber.
The tension grows.
And so does the ambiguity.
This is a realistic comet.
Nothing in the picture compels us to see it as a symbol.

It took me some time to proceed further with this
engraving even with all these clues,
because I could not figure out whether
D\"urer's Gateway to Heaven was
mathematical or religious in nature.
This is a good example of the kind of projection
to avoid in such a study.
First of all, D\"urer  does not separate science and faith.
On the right they entwine in the building and ladder
complex.
On the left,
there is an unnaturally 
straight line from the comet or the light of God
through the Intellectual World of Mathematical Philosophy
to the heart of the Elemental World of
Natural Philosophy.

The Celestial World is not on line, however.
It is old and damaged and out of service.
The Dog remains, idle.
D\"urer graphically reduced the three Worlds to two
in this engraving.
As Leonardo had explicitly said,
the terrestrial and celestial worlds are run
by the same laws,
and are actually one.
The Hermetic doctrine was, ``As above, so below''
but D\"urer -- and possibly Humanist philosophy -- 
dissolved
the absolute separation  between above and below.
The new astronomy saw Earth and stars 
as made of similar stuff governed
by similar laws.

\subsection{Hexagram} %051015

D\"urer was a descriptive geometer, 
practiced in constructing views of bodies from all sides
and in reconstructing a body from its views.
I therefore looked at the Octahedron from all sides, searching for 
his  meaning.
He likely did the same, for he invented the representation of such
polyhedra as nets of polygons \cite{KHULUSI}.

$\bullet$
The ground plan of the Octahedron
 viewed  as a truncated rhomboid  from a certain
 viewpoint
is a Shield of David framed in a hexagon.

The top and 
bottom triangles of the Octahedron project 
into the two crossed triangles of the hexagram. 
Perhaps D\"urer truncated the cube, rather than some other regular solid,
in order to create this hexagram.
It would be anachronistic, however, 
to call this hexagram a Jewish star
and to infer any philo-Semitism. 
Centuries earlier
the Khazars of the Crimea  had adopted the hexagram for their flag when 
they adopted Judaism, 
but in 1514 Nuremberg the hexagram was still mainly a magical device, 
an 
amulet and talisman, possibly referring to the Hebrews of the Bible 
but not to contemporary Jews.\cite{GOLDMAN}.
The Shield of David was called that because it was supposed to
shield its bearer from evil spirits, and 
it was more often found in churches than synagogues. 
A two-footed
hexagram like D{\"u}rer's ground plan is seen on a German altar 
of D{\"u}rer's time. 
Some said the the Seal of Solomon was the hexagram, and others 
the pentagram. 
Traditions firmed later that gave  Solomon's star five 
points to David's six, and stood Solomon's on two points but
 David's on one. 
Not long after 
D{\"u}rer's death the Jewish community of Prague was 
granted the privilege of a flag and chose the 
hexagram, 
already
associated with the Khazars,
and this continued its evolution from magic charm to Judaic symbol. 
D\"urer's use was likely a Hebraism, not philo-Semitism.

Hebraisms  
are  conspicuous in the works of D\"urer and Agrippa.
D\"urer
admired Jerome above the other founding fathers for Jerome's
greater familiarity with the Hebrew Bible
\cite{PANOFSKY}.
In his several drawings of Jerome studying Latin, 
Greek, and Hebrew Bibles, 
D\"urer invariably placed the Hebrew Bible above the others, and
higher is holier.
As his stereotype of First Prophet corroborates, the Hexagram in
the circumcised rhomboid was not philo-Semitism
but  Hebraicism.
It
 connects  the Intellectual Sphere, 
 the Heaven of Humanism, 
 represented by the resulting
 Octahedron, with
the Hebrew Bible,
as Agrippa did.

\begin{figure}
%\vspace{6in}
\begin{center}[In preperation]

$$
\begin{picture}(130,150)
\put(35,115){\line(1,0){60}}
\put(5,65){\line(3,5){30}}
\put(5,65){\line(3,-5){30}}
\put(35,15){\line(1,0){60}}
\put(95,15){\line(3,5){30}}
\put(95,115){\line(3,-5){30}}

\put(5,65){\line(1,0){27}}
\put(125,65){\line(-1,0){27}}
\put(35,15){\line(3,5){13}}
\put(35,115){\line(3,-5){13}}
\put(95,115){\line(-3,-5){13}}
\put(95,15){\line(-3, 5){13}}

\put(48,36){\line(0,1){58}}
\put(32,65){\line(5,3){50}}
\put(32,65){\line(5,-3){50}}

\put(82,36){\line(0,1){58}}
\put(98,65){\line(-5,3){50}}
\put(98,65){\line(-5,-3){50}}

\end{picture}
$$
\caption{The Hexagram in \MI}
\label{fig:0HEXAGRAM}
\end{center}
\end{figure}

\subsection{Bellringer} %050508

D\"urer uses the vertical dimension on his plate in a traditional way. 
Higher is holier.
Jacob's angels get to heaven by ladder.
Angel and Boy have 
equal divinity because they have equal altitude. 
The tools and nails of artisans litter the ground because they are mundane,
while the instruments of science and mathematics hang nearer to the sky because
they are holier.

The horizontal dimension 
is cut in half by the left-most wall  of the House.
The cut separates
the mathematical Gateway II to Heaven on the left from
the theological  Gateway III on the right,
and 
the Rational from the Contemplative Faculties.
Despite first appearances, 
all three Worlds of the cosmos
and all four actors in the drama
are to the left 
of the dividing wall,
and the House, the Ladder, the Comet, 
and most of the Moonbow are on the right with the 
Hourglass, Sundial, Bell, and Magic Square.

The division of the space of the scene by the median plane seems meaningful and
non-random enough
to be  D\"urer's intention.
The extreme point of view of the artist masks this clean cut
by distributing elements of both halves of space
on both halves of the paper,
but all the Biblical and Intellectual elements
are on the right-hand side
of the median wall,
and all the tools and nails are on the left-hand side.

The weighing scales are ambiguously located,
hanging on the dividing wall itself.
This detail fits our interpretation.
Space and time were immaterial
and therefore considered spiritual.
Mass --- ``body''  is the root meaning ---
was material, not geometrical and therefore not spiritual;
it was not defined in terms of space and time
until four centuries later,
by  Albert Einstein.
For D\"urer mass belonged to a lower World than time, 
and probably to the natural philosopher, not the theological.
I suggest that this is why the scales are to the left of the House of God but not the hourglass or bell.

The crucible is no exception.
If it were an instrument of alchemical science,
as I thought at first,  it would be elevated, 
but it is on the stone floor,
so  it is merely a tool of the goldsmith.
Only perspective puts it high on the paper, above all other tools; 
perspective and filial feelings.
This is further indication
 that Doorly read this correctly.
This crucible is not an alchemical instrument 
but a remembrance of D\"urer's goldsmith father,
who died in 1502.
Its position to the far left
gives us one point on the time axis.
The House of God puts Biblical times on the 
right.  

The Octahedron is as opaque as a stone.
D\"urer shows it
 hiding the divine light of Heaven from
the inhabitants of the Elemental World,
the Globe.
The opacity of the Octahedron 
 in the left-hand side 
of the picture 
is the counterpart
to the doorlessness
of the House
on the right-hand side.
Neither mathematical nor theological philosophy
can bring us knowledge of absolute  beauty-truth.

The division of the picture by time 
agrees with its  division by Gateways,
since D\"urer took the occult philosophy
represented by the Chimera on the left
to be later than the theological philosophy and
Scripture represented by the House on the right.
This probably fits his personal experience of them.

The near eye of the Boy  lies on  the center line of the paper,
making him the center of attention,
but he too is left of the median plane.
D\"urer the artist-scientist
takes Agrippa's Gateway I to Heaven,
not his Gateways 2 and 3.

Perspective provides 
a third dimension 
of depth into the scene along the line of sight.
D\"urer seems to use this expressively too.
On the right side our view ends in the House
and on the left there is an infinite vista.
It seems 
that perspective depth in the picture is the familiar 
metaphor  for depth of vision.

All the mortal creatures in the engraving are on the left-hand
 side
of the median plane. 
The only one on the right-hand side is the Bellringer,
off stage, invisible.
There can be little doubt about who is
patiently, patiently
holding the bell-rope,
conducting the Harmony of the Spheres,
and is not to be depicted.
Perhaps 
the engraving was intended for people
who would
prefer this non-depiction of God to
the pagan
representation 
by Michelangelo in the Cistine Chapel,
and perhaps find it
more awesome.

\subsection{Melancholy} %050408

The light and the bow and even the destruction of the 
city of evil are good news, cause for joy,
and
yet  the picture is set in the gloom of night.
D\"urer has even transformed 
the rainbow of the common understanding into a moonbow.
 The comet illuminates nothing.
 Only the moon behind us lights the scene and creates
 the shadows and  the moonbow.
 What might have been a sunny prophesy 
for Leonardo  is  dark as night for Albrecht D\"urer.
 
While Leonardo and D\"urer were both mathematically ambitious,
they differed importantly in their outlooks.
Leonardo wrote confidently of 
``a complete knowledge of all the parts,
which, when combined, compose the totality of the thing which ought
to be loved.'' 
Leonardo was the rational optimist.
D\"urer  on the contrary said shortly before this engraving,
``But what absolute beauty is I know not. 
Nobody knows it but God.''  
The darkness 
and the multiple ambiguities  
suggest  that after all D\"urer 
did not expect humanity to 
attain the supreme mathesis that optimists
like Ramon Llull had
expected and even claimed to have attained;
and that he felt this as a  bereavement.
He expressed
this pessimism in the engraving atmospherically,
by  omitting the door of the House of God,
and 
by numerous skillful demonstrations of 
ambivalence and relativity.

The most unambiguous aspect of this engraving is 
its ambiguity.
D\"urer tells us clearly that we cannot know
anything clearly.
The Octahedron  is a truncated rhomboid with vertical axis, 
or a slab tilted toward the horizon.
There are four subliminal faces  in the Octahedron, or none.
The city is to be destroyed, or not.
The Angel is  sad, or smiling;
the Chimera is good or evil.
The house has windows, or not.
The melancholy is sadness inspired by the inaccessibility
of the absolute, or creative frenzy inspired
by the influence of Saturn,
or because the Celestial World is off its axis.
What we see depends on us as well as what we look at.
The elements of this engraving that I have 
discussed here are deliberately ambiguous
in  meaning or form or both.

What can we believe if we cannot believe what we see?
This questioning
occurs famously in the First Meditation of Descartes (1596-1650),
who cites optical illusions as a reason for his 
method of  universal doubt.
\MI\, anticipates Cartesian doubt.
And its response has a mathematical element that anticipates
the {\em Mathesis Universalis} of Descartes,
but already corrects it by
accepting the limitations to human knowledge.

The ideal of a complete mathematical theory of beauty 
lies
 on the same long line of distinguished fantasies of 
 mathematical wisdom 
as the number mysticism of Pythagoras and Plato, 
the {\em Ars Magna} of Ramon Llull
(whom Agrippa studied) and 
Giordano Bruno (who studied Llull and Agrippa), the vision of 
{\em Mathesis Universalis}
 that Descartes 
and Leibniz shared, 
and the {\em Ars Combinatorix} of Leibniz.
D\"urer does not deny the existence of absolute beauty
but  despairs of knowing it.
The Boy withdraws from both
theology and  astrology
and observes Nature.

Years after 1514, D\"urer like Agrippa
  explicitly abandoned
  the  search for absolute truth and beauty
as futile and hopeless, crying out,
\begin{quote}
``The lie is in our understanding, 
and darkness is so firmly intrenched 
in our mind that even our groping will fail"   
[Panofsky 1971].
\end{quote}
In 1517 D\"urer was still optimistic enough
to join with Luther,
but the engraving already expresses despair in 1514,
so at first I doubted the dating of the engraving.
In fact the date 1514 for the engraving seems reliable, since
several studies for this engraving also bear that date.
Rather,  I infer that D\"urer expressed his pessimism
about attaining absolute truth in his engraving
years before he put it into writing.
Because he was who he was,
first he drew his thought, then he wrote it.
There is no need to doubt the dating of the engraving.

The night of the engraving
is our benighted ignorance.
Its darkness  is the darkness ``firmly intrenched in our mind.''
The mystery of the Octahedron, crafted to be insoluble,
is one of many metaphors in the engraving
 for our inability to see the Absolute  Truth
from our limited perspective.
His chosen Gateway in Heaven is Natural Philosophy,
in spite of its limitations.
In acknowledging the limits of human knowledge,
however,
D\"urer took a step  toward modernity
beyond the more optimistic Leonardo.

The posthumous
publications of D\"urer 
on measurement and on human proportions
show that he never quit
working toward a mathematical 
theory of beauty
and truth
based on measurement
rather than abstract speculation,
but it became more relativistic,
to the degree of
including a variable horizontal scale-factor
transformation
in order to represent human beauty
of the lean, average, or plump variety
as the reference frame of the artist requires.

Agrippa inverted the primary 
meaning of Melancholia just as Horapollo
inverted the social status of the dog, the bat, and the derpent,
making the lowest the highest.
By using the ephemeral 
and counterintuitive
languages of Agrippa and Horapollo,
D\"urer  defies our ordinary rules of interpretation,
making the work surreal but also
contributing to its remarkable endurance.
This melancholia
does not lead to death 
but to its transcendent opposite, a form of 
immortality,
exhibited by the engraving itself, 
vibrant after five centuries.
Actually no one in the engraving is merely sad.
The overall darkness of the engraving
expresses  Agrippan melancholia, not depression but
creative ecstasy,
surely not just of the Boy but also
of the Bellringer Pantocrator.

I find no such quaternities of subliminal faces  in other
works of D\"urer.
The fool of Frau Venus is single, not multi-faced.

D\"urer records that when he made a
present of   \MI\,
he sometimes accompanied it with  {\em Jerome},
but never with the {\em Knight}\/.
This pairing makes sense if the recipients were Protestants.
For obvious reasons,
Protestants
could venerate Jerome but not Erasmus.
Jerome had criticized clerical corruption
to the point of endangering himself,
was a celebrated Hebraist, and
was a close friend of Origen of Antioch, who
is said to have held the Universalist
doctrine,
anathema to the Roman Church,
that ultimately everyone is saved.
Erasmus, on the other hand,
 remained loyal to the post-Nicean Roman Church
 and believed in the value of astrology as a hermeneutic art \cite{BROSSEDER}.
This theory requires only that D\"urer gave
the  prints  in question mostly to  like-minded people.

\subsection{Flourish}
\label{sec:FLOURISH}

A certain Flourish separates the two words 
``MELENCOLIA" and ``I''
of the banderole,
a stylized S that I render as \S.

In 1604 the Flemish engraver Jan Wierix
made a version that is usually called a copy.
He omitted some subliminal elements altogether,
modified others into fiendish masks,
converted the entries in the D\"urer Table to mere numerals,
and deleted the  significant {\S}{}, 
apparently creating a bowdlerized version fit for a pious Protestant market.

$\bullet$  The same flourish  appears in reverse following the date in D\"urer's 1504 ``Adam and Eve".

This makes it likely that the Flourish in \MI\, is a version  
of the S  that precedes the date of the Knight's dateline. 
%	
%	\begin{figure}
%	\vspace{1in}
%	\begin{center}[In preperation]
%	%\includegraphics[width=2.0in]{0PLAQUE}
%	\caption{Plaque from ``Adam and Eve" (1503). }
%	\label{fig:0PLAQUE}
%	\end{center}
%	\end{figure}
It is reversed in one case so that it always faces what it modifies.

The numbers in the Magic Square are determined
 by its $4\times 4$ size and the date line 1514.  
Once the 
bottom line is chosen the rest is no longer under 
D\"urer's control and
so probably carries no information.
However D\"urer still controls the 
orthography of the numerals.
Actually 
there are eight orthographically deviant
numerals in the Table of each version by D\"urer,
and only one in the Wierix version.
Seeing them is easy now because they
call attention to one another:

$\bullet$ The  1 that stands by itself is brushed by the Angel's wing,
and enlarged and elevated 
relative to all the other 1's, as though blessed by the Angel.

$\bullet$ The 4's are not only  numerals
 but also long solid bodies folded into crosses.

$\bullet$ Both 5's are inverted \cite{BRANCH}.
D\"urer ordinarily makes his 5 as we do today.

$\bullet$ One 5 has circular striations
and a faint image of a  numeral in its background.

$\bullet$ The two 6's are more open and overhanging
than  D\"urer's usual 6's,
more like $\sigma$'s than 6's.

$\bullet$ The 9 is open and curly in \MIO.
Its lower end is sinuous, 
with two bends more than the the usual 9.
D\"urer ordinarily makes his 9 as we do today.

$\bullet$ The 9 in \MII\,
is reversed \cite{BRANCH},
more like a $sigma$  than a 9, and is less sinuous,
but still has a serpentine reverse curve in its tail.

$\bullet$ The 0 in \MII\, is a serpent with its own tail
in its jaws.
The 0 in \MIO\, has less detail.

D\"urer's 
precision suggests that these small variations are signicant
and
require interpretation.

When an S appears before a date in D\"urer's work,
it
stands for {\em Salus}\/, or
salvation \cite{PANOFSKY}.
D\"urer  prefixed  the dates of the {\em Knight} and of the
1504 ``Adam and Eve"
with S's
as if to say ``In the year of God 1514'' or ``1504".
The S also occurs in reversed form  as the flourish following the date  in``Adam and Eve''.

The Greek alphabet has two forms of our S, the medial (and initial) form
$\sigma$ and the final form $\varsigma$.
Two forms of lower-case S occur in early English manuscripts and printing. 
Today the medial form has disappeared and the final form 
has taken over, but
both forms are still in common use in D\"urer's time and
both  occur in D\"urer's own writing in (say) the banderole of
his drawing  {\em Orfe der erst Pusaner}\/.  
The prefix to the date of {\em Adam and Eve} is  S.
The variant forms of the numerals 5 and 6  
in the Magic Square can be read respectively
as forms of  final S and medial S (approximately, $\sigma$).
One has the eyes and tongue of a Serpent profile,
another shows a Serpent's face head-on,
so each refers to God twice, once
as abbreviations for ``Salus" and once as the Serpent ideogram for God
found in  the {\em Hieroglyphica}\/.
The two variant 5's and the Flourish in the banderole
are instances of the medial S.
The two variant 6's  are $\sigma$'s, which are also medial S's.
D\"urer's  form of S before the date 1504 in {\em Adam and Eve} is the one most convincingly similar to the 6's in the Square.
The variant 9 is a reflected form of the  final S, and deserves special 
attention. 
The otherwise mysterious variant of 9 after the 
name ``Albert" in  {\em Adam and Eve} 
may be seen as 
 the medial s in the lower left-hand corner
 of the plaque, overturned.

I propose to read all these occurrences of S
as {\em salus} or salvation,  like the Knight's S,
and even as ``God",
especially when they are serpents as well. 
Their meanings are further modified by context.
The 9 following Albrecht's name in {\em Adam and Eve}
would then be another claim of personal divinity.

The blurred background of the 5 in the Magic Square 
recalls the vestigial image of a second hoof
below one of the raised hooves in the {\em Knight}\/.
I suggest that both of these vestigial images 
have the same function as the 
blurred hand of one of the spinners
in the Velasquez painting {\em Las MORE}:
They are pioneering attempts to show motion in an otherwise static medium.
The duplicated hoof is the one most rapidly moving; the duplicated 5 is
evidently spinning in its box, as shown also by the circular
trails behind it.

The 1  in D\"urer's Table 
and the ``I'' in the Motto are likely synonymous, 
the Flourish in the Motto serving like the Angel's wing
in the Table
to show that the 1 is holy.
I take  these 1's to refers to Agrippa's Gateway I,
Natural Philosophy.
This was capital heresy.

We thus cover all 10 graphic 
anomalies in ``Melencolia I'', plus
one in the ``Knight" and two in ``Adam and Eve''
with one hypothesis:
They depict God.
Combined with  
the Wave
 and the twenty subliminal faces, 
they make ``Melencolia I'' densely pro-science, pro-art, 
 and so anti-Church.

As though to verify this interpretation,
Jan Wierix  in his 1605 version of \MI
 systematically modified
D\"urer's work:

$\bullet$  Wierix
eliminated the ambiguous Wave.

$\bullet$ He deleted the Flourish.

$\bullet$ He reduced the magnification and elevation of the solitary 1
in the Magic Square.

$\bullet$ He flattened the 4's and eliminated their folds, 
so that they are no longer solid crosses but mere numerals.

$\bullet$ He corrected the 5 within the date in the Magic Square.

$\bullet$ He corrected the 6 not touching the date
in the Magic Square.

$\bullet$  He retained the variant 5 touching the date
in the Magic Square like the S of the Knight.

$\bullet$  He retained the variant  6  prefixed to the date
in the Magic Square like the S of the Knight.

$\bullet$ He retained the variant 9 
before the date in the Magic Square.

 I see no other significant changes.

Wierix thus stripped the engraving of  six  heretical
graphic elements
and retained two  conforming elements.
This suggests that he
wished to make a cleansed version suitable 
for Flanders.
If we suppose that he could have eliminated 
each of these nine features as likely as not,
the probability of doing all this by chance is 
$2^{-9}=1/1024$\/.

Wierix also modified the subliminal faces.
He almost eliminated all four Ghosts,
replaced the quaternity of Friends by one cadaverous face,
and eliminated First Prophet, 
the worm and skull-cap of Second Prophet, 
and First Fool.
No quaternity or trinity remained when he was through.
He also eliminated the ambiguous smile of the Angel.
His Angel is unmistakably serious, even sad.
%	
%	\begin{figure}
%	\begin{center}[In preperation]

%	\vspace{1in}
%	\caption{\MI\, of Jan Wierix}
%	\label{fig:WIERIX}
%	\end{center}
%	\end{figure}

I conclude that probably some form of my interpretation 
of these nine graphic elements
was still alive in 1605 and known to Wierix;
I have merely revived it, not invented it.

These interpretations of the wandering S's and crosses
make sense only if
there were viewers who would understand them.
Since the doctrines of ``Melencolia I'' 
are part
of the Humanist tradition,
most likely  the initiates 
were  Humanists,
especially D\"urer's 
best friend, Willibald Pirckheimer.

\subsection{Serpents}

After writing the above
I realized that I had noticed but 
not understood the sinuousity
of the 9 in \MIO
and examined the figure more closely,
 to find that it is 
a fully realized serpent.

$\bullet$ 
The top of the 9 in the Table has a Serpent's head  
curled inside the loop of its body, 
with two eyes
looking directly at us,  a mouth-line,
and a forked tongue.

In \MIO\, I must cock my head about $30^{\circ}$
to the left to best see this face with its two eye-dots.
In \MII the 9 is reversed,
the serpent's face is upright,
and the eyes have grown from miniscule dots into discs,
all serving to make the Serpent more conspicuous.

$\bullet$ 
The numerals 6 in the D\"urer Table are also serpents, with head and forked tongue
at their lower end.
These
are the most minute of the  subliminal structures.
The forked tongues are conveyed by
two scratches of the burin.

$\bullet$ 
The numeral 0 in the D\"urer Table of \MI\,
is a Serpent eating its tail,
the Ourobourous.
The eyes are drawn only in \MII.

The Serpent in the 9 seemed to refute my then current interpretation.
The S that I had deduced was a sign of the Highest
was the lowest of the low.
But within the hour  I remembered that I had 
experienced this inversion
before,
first with the Dog and then with the Chimera.
I went back to Horapollo and looked up 
``Snake.'' 
Nothing.

Then ``Serpent:''
\begin{quote}
{They symbolize the Almighty by the perfect animal,
again drawing a complete serpent.
Thus among them that which pervades the whole cosmos
is Spirit.  (\cite{HORAPOLLO}, Book I, Symbol 64)}
\end{quote}
So the Serpent in the Table does not wreck the interpretation
 but supports it.
 
The Serpent among the 
three Prophets in \MIO\, presumably has the same meaning.
In \MIO,
what I first saw as a worm extrudes from the right eye
of Second Prophet.
It is not especially different in shape from
what I have read as heads of serpents in the Magic Square,
but
 I read it as a worm
and understood D\"urer to say that Jesus was dead.
 This led me to read Unitarianism into the picture.
Coming across \MII\,  was like being corrected by D\"urer
in conversation.
Perhaps D\"urer had become aware of just
this misunderstanding and used
\MII\, to make his meaning clearer.
In  \MII\, he  enlarged 
 and elaborated the creature that I had seen as a worm in \MII\,
 into what seems most like the head of an Eagle,
  still apparently emerging from
Second Prophet.
The eagle is another symbol for God in {\em Hieroglyphica}\/

Then  
there are no indications in \MI\ that D\"urer differed
with Christian orthodoxy on the divinity of Jesus;
quite the contrary,
it is a Humanist manifesto, not a Unitarian one.
The first known Unitarian publication came more than a decade 
after this engraving.

This ended my search for Fourth Prophet.
 There are only the three Prophets,  presented as 
 aspects  of  God,
 represented by the Eagle in \MII.
 In this theory,
 D\"urer hides the Bellringer in obedience
 to the commandment about graven images
but show the Serpents and the Eagle
because they are not images of God
but merely glyphs.

We see that the Magic Square is riddled with defects, such as
 inverted, backwards, or otherwise
distorted numerals.
I suggest that  these imperfections were meant to disqualify this
engraving for the talisman trade, which still flourishes.
Not only did D\"urer disdain the magic arts,
it would have been dangerous to be suspected of practicing them.
Most of the distortions inject symbols of God into the Magic Square,
as though to fend off accusations of devil worship.

\subsection{Summation}

There are several main 
ideas beneath the surface of this engraving:

1. 
The Motto 
``MELENCOLIA{\S}I", in its overt form,
and the general gloom of the picture,
refer to the melancholy of 
those who strive in vain
to define or create absolute truth and 
beauty.
This limitation of  knowledge  
inspires a melancholia in the Boy that 
is not  depression but a frenzied and sanctified
creativity.

2. Unscrambled, the Motto becomes LIMEN CAELO I,
probably referring to 
Agrippa's Gateway to Heaven I, 
Natural Philosophy.
As Gateways to Heaven,  Natural Philosophy, based on the observation and portrayal of nature in
mathematical terms, is on one level with
Theological Philosophy, 
based on the Holy Scriptures.
The creative thinkers of the early sixteenth century  
did not separate art and science as sharply as those of the twenty-first.
When D\"urer sanctifies artistic depiction he also sanctifies scientific observation.
True, Natural Philosophy is not a sure Gateway to Heaven,
as shown by the various perspectival ambiguities.
But neither is Theological Philosophy, as shown 
by the three warring sects beneath her dress.
And Mathematical Philosophy, meaning  astrology,
is not even in the running:
 D\"urer discards the Celestial Sphere, 
 the middle world of the Neo-Platonic triple cosmos,
together with astrology, the Mathematical Philosophy of Agrippa.

This is
a Humanist manifesto of the impending 
Reformation,  the scientific revolution,
and Natural Philosophy.

D\"urer anticipates Descartes' universal doubt 
based on the existence of optical illusions, 
and also Descartes' two-fold world
of matter (the Elemental World of {\em res extensa}) and mind
 (the Intellectual World of {\em res extensa}),
but he is already beyond Descartes in 
recognizing
 that our mathematical models can never exactly match reality,
that because perception is the source of all our knowledge,
our knowledge is relative, not absolute.

We read these messages
in the grand overall structure, the fine details,
the ambiguous
perspective projections,
the quadrivalent optical illusions,
the orthographic variants,
and the anagram of this engraving.

Some of these messages  were capital crimes 
in the time and place of D\"urer
and account for the concealments 
in the engraving.
 Yet the  concealments  
are not mere self-protection,
as I assumed at first.
The multiple ambiguity does not merely
cloak the message, it {\em is} the message.
\begin{quote}
``The lie is in our understanding, 
and darkness is so firmly intrenched 
in our mind that even our groping will fail."   
\end{quote}
For D\"urer and \MI,
the world still has a unique meaning,
but God alone knows what it is.
\MI\,  with its pervasive ambiguity,
is a postmodern work of art done in the 16th century.

\section{POSTSCRIPT}
I disclose here some of my own
views on these topics of D\"urer.

Some of the Neo-Platonic cosmology used  in \MI\,
still lingers on today as
the crippling belief 
that Nature is a mathematical system.
I call this
the mathetic fallacy,
the mathematical 
counterpart of the pathetic fallacy.
It still persists even though today
 we understand better the evolutionary
process that produced and still produces mathematics
and physics.
The monism that pervades
\MI\,
survives in physics today 
as the passion for unity and simplicity, and
competes
with the hunger for richness and complexity,
of which the engraving itself is a specimen.
The  limits to knowledge  illustrated in \MI\,
agree broadly with some major thinkers of the time.
But today's skepticism cuts even deeper.
We suspend belief not merely in our ability 
to perceive absolute reality but 
in the very existence of such an absolute.
Our predicament is less melancholy than D\"urer's,
since we are not deprived of anything that ever existed,
and more hopeful, since the reduction in our certainty 
comes with 
an increase in our potentiality.

The Renaissance doctrine
that all knowledge comes through the senses
oversimplified the story.
It undervalues hereditary
 knowledge, 
such as the vital procedural knowledge of how to breathe and swallow
and learn to speak.
It makes us seem
neutral
observers when  actually we 
have vested interests. 
It makes us seem passive observers,
when  actually we 
act as much as we react when we study nature,
and there are many nerves running from the brain to the eye
as well as the other way.
Only prior knowledge enables us
to convert what comes through the senses
into new knowledge.
Descriptive geometry is an important beginning
but 
the Boy still has to learn 
to value experiments.
We need to know what works,
not just what is.

D\"urer, Kepler,  and Einstein  wanted
to know the creative mind of God,
at least metaphorically:

\begin{quote}
``I want to know how God created this world.  I am not interested in this or that phenomenon, in the spectrum of this or that element. I want to know his thoughts; the rest are details.'' \hfill   (Attributed to Albert Einstein)
\end{quote}

This metaphor is an insidious form of the mathetic virus
and the dogma of intelligent design.
It led Kepler to
his Platonic-solid model of the Solar System,
a thoroughly mistaken theory,
and blocked him from  appreciating
the beauty of
the three laws that he found
along
a more empirical path
and  that are still named after him.

When I began my own research
 I took it for granted that it had three stages:
I would first find a theory in which I could at least potentially believe, 
then compute its consequences, test it against experimental data, 
and return to stage 1 for an improved version.
After about forty years I could not help noticing 
that I was still in the first stage.
It may be, I begin to surmise,  
that physics is not the project of replacing one faith by another,
but replaces the method of faith by another method, 
the method of experiment.
I have still not found a theory I can believe in, but now I see this as 
a vital sign, 
a source of joy, not melancholy.

Some physicists still cherish the concept of a single
 eternal law of nature
or an all-encompassing theory,
and the quest for one has stimulated much important
and fruitful work,
but the concept derives
from misplaced faith like Kepler's,
not  from experiments
like Galileo's, Rutherford's, and Hubble's.
The concept continues to divert 
us from the experimental method today but
I am not sure it is even self-consistent:
Must the final theory describe itself?
Must it predict what experiments we are going to choose to do
to test it?

There
 is no mathematics, no poetry, no physics
in raw Nature. 
They are part of
 culture,
a small highly ordered part of Nature.
We can find the meaning in Nature not by looking
through a microscope or a telescope 
but only by looking out the window, at people and at
our own reflection in the glass,
for we create it.
Nature is no theory but as the name itself reminds us
an evolutionary process that births us and the theories
we secrete.
Our theories are
but millstones 
that
continually fall off their axles
and have to be replaced.

\subsection{Acknowledgments} %050508

The study would have been impossible 
without the works  on ``MELENCOLIA I"\/ by Erwin Panofsky 
and by Francis Yates  and  the studies of Agrippa by
Judit Gell\'erd and by Charles Nauert.
The heraldry dictionary of 
Cecil Wade helped me to read D{\"u}rer's coat-of-arms.
I thank 
Dr. Basimah Khulusi and Prof. Bill Branch for
their interpretations of several elements of \MI,
some of which I ultimately accepted.
Khulrusi pointed out the reference to Jesus in the four nails of the engraving,
and made a convincing model of the 
Octahedron that established the angle 89.5$^\circ$,
among numerous other contributions;
Branch pointed out that the numeral 5
in the Magic Square 
was upside down and that the 9 was backwards,
and brought up the Ouroubouros.
I noted some  numerology attributed to D\"urer 
on the Web site of the Aiwaz Gallery though finally I did not accept it.
My thanks go to Roy Skodnick 
for introducing me to the work of Yates; 
to Prof. Heinrich Saller 
for the print of ``MELENCOLIA I" that triggered this process; 
to Rabbi Mario Karpuj for reminding me
that Jacob's ladder is the Biblical gate of heaven; 
to Prof. Shalom Goldman for stimulating, informative, and 
encouraging discussions, the reference to Yates on D{\"u}rer, 
and the meaning of 
the Magen David in the Renaissance; 
to Danny Lunsford for the reference to http://meteoros.de/halo/halo1.htm;
to Carla Singer 
for lending me her expertise in art history, including the reference to Mantegna; 
to Aria Ritz Finkelstein for the reference to Velasquez; 
and to Shlomit Ritz Finkelstein, for lively discussions, 
help with Torah and Descartes, and 
useful expository suggestions. 
A first stage of this study  was published in 
{\em The St. Ann's Review}, whose reviewer improved the paper
\cite{FINKELSTEIN}.
A second stage was submitted for publication with
Dr. Khulusi \cite{FINKELSTEINKHULUSI}.
A new version with illustrations is in preparation.

% {\addtolength{\parskip}{-12pt}} %Put at top of toc

%\listoffigures
%\tableofcontents
\end{document}